\documentclass[12pt, draftclsnofoot, onecolumn]{IEEEtran}
\IEEEoverridecommandlockouts
\usepackage{amsmath,cite,bm,amsfonts,amssymb,graphicx,color,mathtools,setspace,comment,multirow,xfrac}

\usepackage{epstopdf}
\usepackage{footnote}
\usepackage{multirow}
\usepackage{lscape}
\usepackage{graphicx}
\usepackage{float}
\usepackage{amsthm}
\usepackage{alltt}
\usepackage{placeins}

\newcommand{\Pt}{P_{\textrm{t}}}

\newcommand{\bh}{\mathbf{h}}
\newcommand{\bH}{\text{H}}

\newcommand{\ba}{\mathbf{a}}

\newtheorem*{result1}{{Theorem 1}}
\newtheorem*{result2}{{Theorem 2}}
\newtheorem*{result3}{{Result}}

\newcommand{\squeeeeze}{\hspace{-0.04cm}}

\newcommand{\myincludegraphics}
{\includegraphics[trim=0.7cm 0cm 1.3cm 0.8cm, clip=true, width=0.5\columnwidth]}

\begin{document}

\bstctlcite{IEEEexample:BSTcontrol} 

\title{Massive MIMO Asymptotics for Ray-Based Propagation Channels
\thanks{Parts of this work were presented at the IEEE International Conference on Communications (ICC) 2019 \cite{lishua}.}	
\thanks{The work of S. Li was supported by China Scholarship Council. The work of M. Matthaiou was supported by EPSRC, UK, under grant EP/P000673/1.}
\thanks{ S. Li and P. A. Dmochowski are with the School of Engineering and Computer Science, Victoria
	University of Wellington, PO Box 600 Wellington 6140, New Zealand (e-mail: \{lishua, pawel.dmochowski\}@ecs.vuw.ac.nz).}
\thanks{P. J. Smith is with the School of Mathematics and Statistics, Victoria
	University of Wellington, PO Box 600 Wellington 6140, New Zealand (e-mail: peter.smith@ecs.vuw.ac.nz).}
\thanks{H. Tataria is with the Department of Electrical and Information Technology, Lund University, Lund, Sweden (e-mail: harsh.tataria@eit.ith.se).}
\thanks{M. Matthaiou is with the Institute of Electronics, Communications and Information Technology (ECIT),
	Queen’s University Belfast, Belfast, U.K. (e-mail: m.matthaiou@qub.ac.uk).}
\thanks{J. Yin is with the College of Underwater Acoustic Engineering, Harbin Engineering University, Harbin 150001, P.R.China (e-mail: yinjingwei@hrbeu.edu.cn).}
%
%
%
%
%
%
}

		\author{\color{black}
	Shuang~Li,~\IEEEmembership{Student Member,~IEEE,}
	Peter~Smith,~\IEEEmembership{Fellow,~IEEE,}	
	Pawel~Dmochowski,~\IEEEmembership{Senior Member,~IEEE,}
	Harsh Tataria,~\IEEEmembership{Member,~IEEE,}
	Michail Matthaiou,~\IEEEmembership{Senior Member,~IEEE,}
    and Jingwei Yin, ~\IEEEmembership{Member,~IEEE}}
 \color{black}

\markboth{IEEE Transactions on Wireless Communications,~Vol.~XX, No.~XX, XXX~2019}
{}

\maketitle

\vspace*{-0.5cm}
\begin{abstract}

	Favorable propagation (FP) and channel hardening (CH) are desired
	properties in massive multiple-input multiple-output (MIMO)
	systems. To date, these
	properties have primarily been analyzed for classical \textit{statistical}
	channel models, or \textit{ray-based} models with very specific angular
	parameters and distributions. This paper presents a thorough mathematical analysis of
	the asymptotic system behavior for ray-based channels with
	\textit{arbitrary} ray distributions, and considers \textit{two} types of antenna array structures at the cellular base station: a uniform linear array (ULA) and a uniform planar array (UPA). In addition to FP and channel
	hardening, we analyze the \textit{large system potential} (LSP) which
	measures the asymptotic ratio of the expected power in the desired channel to the expected total interference power when both the antenna and user numbers
	grow. LSP is said to hold when this ratio converges to a positive constant.
	The results demonstrate that while FP is
	guaranteed in ray-based channels, CH may or may not
	occur depending on the nature of the model. Furthermore, we
	demonstrate that LSP will not normally hold as the expected interference power
	grows logarithmically for both ULAs and UPAs relative to the power in the desired channel as the system size
	increases. Nevertheless, we identify some fundamental and attractive
	properties of massive MIMO in this limiting regime. 
\end{abstract}



\IEEEpeerreviewmaketitle

\section{Introduction}


Two key principles behind the success of massive MIMO are favorable propagation (FP) \cite{ref7,ref51new}, and channel hardening (CH)\cite{ngo2017no}, meaning that  the normalized inter-user interference power converges to zero, and that the normalized power in the desired channel becomes constant. With FP, the use of large numbers of antennas offers an implicit interference reduction mechanism, and enables the use of low complexity signal processing algorithms \cite{ref51new,ref52,rusek2013scaling}.


The bulk of the theoretical work on FP and CH has utilized classical \textit{statistical} channel models. Here, the existence of FP has been demonstrated for channel models of increasing complexity, progressing from  independent and identically distributed (i.i.d.) Rayleigh \cite{ref7,ref9}, pure line-of-sight \cite{ref7,ref9}, correlated Rayleigh \cite{ref3,ref4}, and independent Ricean \cite{ref1} to correlated Ricean channels \cite{matthaiou2019does,ref21}. In parallel, with the theory, channel measurements have demonstrated that a large fraction of the theoretical gains due to FP can be obtained \cite{ref10,hoydis2012channel,gao2015massive}. 

This work is now mature, but \textit{incomplete} in the sense that accurate modeling of large dimensional channels requires a strong link to the propagation environment. This is usually obtained through ray-based models which have been extensively validated by measurements and, for this reason, have been adapted in the 3GPP standardization\cite{ref26}. These models better capture the physics of electromagnetic propagation, have a closer link to the array architecture and are widely used irrespective of the frequency band \cite{ref26,sangodoyin2018cluster,ref53}. The physical nature of the ray-based models also has advantages in system performance analysis \textcolor{black}{since the analytical conclusions are based on physical features of the system rather than statistical modeling assumptions}. For example, FP was considered in the recent work\cite{matthaiou2019does} for very general heterogeneous, correlated Ricean channels. This work gives wide ranging FP results, but the inherent nature of these models meant that the conclusions relied on various assumptions concerning the correlation structure, line-of-sight direction, etc. In contrast, we are able to prove FP for ray-based models with the most basic assumptions pertaining to physical phenomena, such as  ray direction distributions. 

Variations of such models have a proliferation of names including directional, spatial and Saleh-Valenzulela (SV) type \cite{saleh1987statistical} channel models. We prefer the phrase \textit{ray-based}, as the main requirement for our work is that the statistical distributions of individual rays can be identified and analyzed.\footnote{We prefer this terminology, since our primary interest is not in identifying distributions of rays across multiple clusters.} This is possible for a wide range of such channels. Important work has begun in this area demonstrating the existence of FP with specific ray-based models for a variety of antenna topologies, such as the uniform linear array (ULA), uniform planar array (UPA), and uniform circular array (UCA) \cite{wu2017favorable}, \cite{ref18}.  However, the majority of this work relies on \textit{two} very special cases for the rays: an arbitrary ray must arrive with an azimuth angle, $\phi$, which satisfies $\phi \sim U[0,2\pi]$ \cite{wu2017favorable,roy2018mimo} or ${\textrm{sin}}\phi \sim U[-1,1]$\cite{ngo2014aspects}. 
FP has also been demonstrated in \cite{zhang2019favorable} for the more complex case where azimuth angles have a uniform central angle and wrapped Gaussian sub rays and the elevation angles are Laplacian. However, with the exception of our work in \cite{lishua}, there are no general results available for arbitrary angular distributions. It is, thus, critical to be able to predict the system performance with practical, and more general distributions, making our extension an important one in the context of the understanding of massive MIMO behavior. Hence, a general analysis of FP for ray-based models with arbitrary ray distributions is almost entirely lacking. Further, while FP is a desirable property for a communication system, it only implies that a finite number of users can be served by increasing the number of antenna elements. We refer to this as \textit{single-sided} massive MIMO\cite{wu2017favorable}. Ideally, as you grow the number of BS antennas you would also serve more users, leading to a system that becomes large both in users and antennas, i.e., \textit{double-sided} massive MIMO, a concept defined in \cite{buzzi2018energy} . Hence, we define large system potential (LSP) as the property that the fundamental ratio which measures the \textcolor{black}{mean} power in the desired channel relative to the total \textcolor{black}{mean} interference power converges to a positive constant as both the number of users ($K$) and the number of antennas ($N$) grow to infinity, with $N/K \rightarrow \alpha$ as $N \rightarrow \infty$. The analysis of LSP for i.i.d. Rayleigh fading \textcolor{black}{can be found in \cite{rusek2013scaling}, although not explicitly defined}. In our conference paper \cite{lishua},  CH, FP, and LSP analysis for ray-based channel models and a basic ULA antenna structure were discussed. Here, we extend our earlier work in \cite{lishua} by also considering a UPA structure. This is an important, yet non trivial, extension as the addition of the elevation component in the ray based channel facilitates much greater accuracy in predicting massive MIMO behavior, something which is rare in the literature. Furthermore, the majority of the ongoing deployments of massive MIMO in the C-band utilize UPAs, in order to leverage the full dimensional nature of the channel, and to maximize the beamforming gain of the system via reciprocity-based processing. The mathematical complexity of such an extension is substantial, since our main aim is to obtain detailed insights into the desirable properties of massive MIMO systems with a model which is more closely in line with practice. We make the following contributions for both ULA and UPA topologies: 
\begin{itemize}
	\item We show that CH may or may not occur depending on the nature of the model.
	\item We show that FP is guaranteed for all models where the ray angles are continuous random variables (as assumed by all models to date). 
	\item For LSP, we derive remarkably simple expressions which relate the asymptotic interference behavior to system size, antenna spacing the ray distribution.  We demonstrate that LSP will \textit{not} normally hold as the \textcolor{black}{mean} interference power grows logarithmically in $N$ relative to the \textcolor{black}{mean} power of the desired user channel as the system size increases.
\item Analytical results are verified via simulation and using \textcolor{black}{exact} closed-form special cases \textcolor{black}{derived} for specific angular distributions.
	\item Despite the lack of LSP, the implications for massive MIMO are excellent. Although the interference eventually dominates the desired channel, the growth is very slow and is further attenuated by practical factors such as the likely propagation environment and the typical array patterns employed. In addition, we prove that trivial scheduling schemes can retain LSP, and thus increase the robustness of massive MIMO performance.
	     
\end{itemize}
\par 
\textbf{Notation.} Boldface lower and upper case symbols denote
vectors and matrices. Complex conjugation and Hermitian transpose operations are denoted by $(\cdot)^*$ and $(\cdot)^\text{H}$, respectively, while $\otimes$ denotes the Kronecker product; $\mathcal{CN}(\textbf{{m}}, \textbf{{R}})$ denotes the circular symmetric complex Gaussian distribution with mean $\textbf{{m}}$ and covariance matrix $\textbf{{R}}$, while $U[a,b]$ denotes a uniform distribution on $[a,b]$;  $\mathbb{E}[\cdot]$ denotes statistical expectation, $\sim$ denotes asymptotic equivalence defined in \cite[p.~15]{abramowitz1964handbook}, $\xrightarrow{\text{a.s.}}$ denotes almost sure convergence, and $\mathcal{F}(\cdot)$ denotes the Fourier transform. The functions $J_0(\cdot)$ and $I_0(\cdot)$ denote the zero order Bessel and modified Bessel functions of the first kind, while $O(\cdot)$ denotes the growth rate of the argument. The Dirichlet kernel is defined as D$_{n}(x)=\dfrac{\text{sin}((n+{1/2})x)}{2\pi\text{sin}(x/2)}$ for some positive integer $n$.

\section{Channel Model and System Metrics} \label{section:System_Description}
We consider an uplink massive MIMO system with  $N$ co-located antennas at one base station (BS) simultaneously serving $K$ single antenna users, where, for now, $N\gg K$. We assume a narrowband flat fading channel model such that the $N\times1$ channel vector for user $i$ can be written as $\textbf{h}_{i}$, and the composite $N\times K$ channel matrix is denoted by $\textbf{H}=[\textbf{h}_{1} \textbf{h}_{2}...\textbf{h}_{K}]$. We assume that the propagation channel is known to both the users and the cellular BS. This is an assumption in this study since we are concerned with analyzing the fundamental properties of a massive MIMO system using ray-based channels. In practice, reciprocity-based beamforming may be used where uplink channel estimates will be used for payload data transmission.  

\subsection{Ray-based Channel Model}
In general, the propagation channel to user $i$ can be described as the superposition of many individual rays possibly arriving in clusters from a set of far-field scatterers. In simple terms, the channel is broken down into $P$ incident rays at the BS.%
\footnote{For ease of notation, we do not specifically itemize clusters, but the $P$ paths include any clustered rays. We note that since we are concerned with angular distributions of individual rays contributing to the channel impulse response, it is not necessary to categorize the channel model in terms of clusters.} Hence, for a ULA we have
\begin{equation}\label{SVchannel}
\bh_i=\sum_{r=1}^P\gamma_{ir}\ba(\phi_{ir}),
\end{equation}
where  $\phi_{ir}$ is the azimuth angle of the $r^\text{th}$ ray, $\gamma_{ir}$ is a complex scaling factor for the magnitude and phase of the ray, and $\ba(\phi_{ir})$ is the $N\times 1$ array steering vector.  
In azimuth, the antenna array broadside is at $\phi_{ir}=0$, and end-fire is $\phi_{ir}=\pm\frac{\pi}{2}$. Common models for the scaling factor include random phase models \cite{ref26}, where $\gamma_{ir}=\sqrt{\beta_{ir}}{\text{exp}(j\Phi_{ir})}$, $\beta_{ir}$ is the power of the ${r}^\text{th}$ ray and $\Phi_{ir}$ are i.i.d.  $U[0,2\pi]$ phase offsets. Hence, $\beta_{i}=\sum_{r=1}^{P}\beta_{ir}$ is the total link gain for user $i$. Also, complex Gaussian models have been proposed in \cite{ref53}, where $\gamma_{ir}=\sqrt{\beta_{ir}}u_{ir}$ and 
$u_{ir}\sim\mathcal{CN}(0,1)$. For both models, we note that $ \mathbb{E}[\gamma_{ir}]=0$, $\mathbb{E}[|\gamma_{ir}|^2]=\beta_{ir}$ and $\mathbb{E}[\gamma^*_{ir}\gamma_{js}]=0$ for all pairs $(i,r) \neq (j,s)$. 
For a ULA with normalized inter-element spacing $d$, measured in wavelengths, the steering vector is given by
\begin{equation}\label{steeringULA} \nonumber
\ba(\phi_{ir})=\left[1, e^{2\pi j d {\textrm{sin}}\phi_{ir}}, e^{2\pi j 2d {\textrm{sin}}\phi_{ir}}, \ldots ,e^{2\pi j (N-1)d {\textrm{sin}}\phi_{ir}}\right]^{\textrm{T}}.
\end{equation}

For a uniform planar array (UPA), the total number of antennas, $N$, is divided into $N_x$ and $N_y$ antennas in the $x$ and $y$ axes with inter-element spacings $d_x$ and $d_y$, respectively. The steering vectors can be represented by the Kronecker product of ($N_x\times{1}$) $\ba_x$  and ($N_y\times{1}$) $\ba_y$, which gives an $N_x{N}_y\times1$ vector,  $\ba(\theta_{ir}, \phi_{ir})=\ba_{irx}\otimes\ba_{iry}$. Note that
\begin{align} \nonumber
\ba_{irx}&=\left[1, e^{2\pi j d_x {\textrm{sin}}\theta_{ir}{\textrm{cos}}\phi_{ir}}, \ldots, e^{2\pi j (N_x-1)d_x {\textrm{sin}}\theta_{ir}{\textrm{cos}}\phi_{ir}}\right]^{\text{T}},
\end{align}
and
\begin{align} \nonumber
\ba_{iry}&=\left[1, e^{2\pi j d_y {\textrm{sin}}\theta_{ir}{\textrm{sin}}\phi_{ir}}, \ldots, e^{2\pi j (N_y-1)d_y {\textrm{sin}}\theta_{ir}{\textrm{sin}}\phi_{ir}}\right]^{\text{T}}.
\end{align}
Note that this definition of the steering vectors follows the notation in \cite{wu2017favorable} where the array is defined  in the $(x,y)$-plane. Hence, $\phi_{ir}$ is the  angle of the $r^\text{th}$ ray for the $i^\text{th}$ user in the $(x,y)$-plane relative to the $x$-axis. The angle $\theta_{ir}$ is the angle of the $r^\text{th}$ ray for the $i^\text{th}$ user measured from the zenith direction ($z$-axis). With this definition,  if the UPA is horizontally oriented then $\phi_{ir}$ is an azimuth angle and  $\theta_{ir}$ is an elevation angle. For vertically located arrays, the more general definition of the angles applies, \textcolor{black}{where $\theta_{ir}$ and $\phi_{ir}$ are defined relative to the z and x axes, respectively.} The $k^\text{th}$ elements of $\ba_{irx}$ and $\ba_{iry}$  are denoted by $\ba_{irx{k}}$ and $\ba_{iry{k}}$, respectively. In practice, each element has an active directional gain pattern, which attenuates the radiated power as a function of the steering direction. In order not to obfuscate the results and deviate  focus from the asymptotic massive MIMO properties,  we assume that each element has an equal gain in all directions (i.e., omni-directional), irrespective of the element location in the array.
\subsection{FP, Channel Hardening and Large System Potential}
Here, FP denotes asymptotic FP  where ${\bh_i^\textrm{H}\bh_j}/{N} \xrightarrow{\text{a.s.}} 0$ as $N\to \infty$\cite{ref7}. Channel hardening refers to the property that ${\bh_i^\textrm{H}\bh_i}/{N} \xrightarrow{\text{a.s.}} \beta_{i}$ as $N\to \infty$, which is equivalent in our case to the definition in \cite{ngo2017no}. Now,  FP and CH imply that the interference from one user to another vanishes relative to the signal power almost surely as $N\rightarrow\infty$. We extend this definition to the asymptotic regime where $N\rightarrow\infty, K\rightarrow\infty$ and $N/K\rightarrow\alpha$ (double-sided massive MIMO). Here, the equivalent question concerns whether the total interference power  to user $i$ dominates the signal power. In order to investigate this behavior, we define 
\begin{equation} \label{SIR}
{\zeta_\textrm{LSP}}=\frac{\mathbb{E}[|\bh_i^\bH\bh_i|^2]}{\sum_{j\ne i}^{K}\mathbb{E}[|\bh_i^\bH\bh_j|^2]}=\frac{\mathbb{E}[|\bh_i^\bH\bh_i/N|^2]}{\sum_{j\ne i}^{K}\mathbb{E}[|\bh_i^\bH\bh_j/N|^2]}.
\end{equation}
Now, $\zeta_\textrm{LSP}$ is a fundamental performance metric, measuring the ratio of the desired mean channel power to the total mean interference power. 
We say that LSP holds if $\zeta_\textrm{LSP}$ converges to a positive constant as $N \to \infty$ and $N/K \rightarrow \alpha$. \textcolor{black}{If CH holds, then $\bh_i^\bH\bh_i/N$  converges to a positive constant and hence,   the numerator of (\ref{SIR}) also converges to a positive constant. Even when CH does not occur (see Sec.~\ref{Channel Hardening}), the expectation in the numerator of (\ref{SIR}) will converge to a constant.}

If CH holds then the numerator of (\ref{SIR}) converges to a positive constant. Hence, LSP depends on the limiting behavior of the denominator of (\ref{SIR}), namely $\mathbb{E}[\eta_{i}]$, the mean of the total normalized interference, where $\eta_i$ is defined as
\begin{equation}
	 \eta_{i}=\sum_{j\ne i}^{K}|\bh_i^\bH\bh_j/N|^2.
\end{equation}
Note that the limiting regime used for  LSP, which supports double-sided massive MIMO, is far more challenging than traditional massive MIMO. In practice, the number of users will never grow without bound but the asymptotics are still useful in identifying the key properties of systems which are large in both $N$ and $K$. 

\section{ULA: Channel Hardening, FP and LSP}\label{sec: fixedSV}
CH, FP and LSP are now considered for ray-based channels for a ULA, and let $\ba_{ir}$ be the steering vector for user $i$, path $r$, $\ba_{ir}=\ba(\phi_{ir})$ and the $k^\text{th}$ element of $\ba_{ir}$ is denoted $\ba_{irk}$. 
\subsection{Channel Hardening}\label{Channel Hardening}
Consider the term, $\bh^{\bH}_i\bh_i/N$, for a ULA. We have
\begin{align} \label{eq:FP ULA}
\dfrac{\bh_{i}^{\bH}\bh_{i}}{N}=\frac{1}{N}\sum_{r=1}^{P}\gamma^{*}_{ir}\ba_{ir}^{\bH}\sum_{s=1}^{P}\gamma_{is}\ba_{is}
=\sum_{r=1}^{P}|\gamma_{ir}|^2+\frac{1}{N}\sum_{r=1, r \neq s}^{P}\sum_{s=1}^{P}\gamma^{*}_{ir}\gamma_{is}\ba_{ir}^{\bH}\ba_{is} 
=X_{i} + E_{i},
\end{align}
where $X_{i}=\sum_{r=1}^{P}|\gamma_{ir}|^2$ is independent of $N$. Thus the limiting value depends entirely on $\lim_{N\to\infty}E_{i}$, which in turn depends on $\lim_{N\to\infty}\ba_{ir}^{\bH}\ba_{is}/{N}$, where $r \neq s$. 
Now,
\begin{align} \label{eq:signal array response vector}
\left\lvert \dfrac{\ba_{ir}^{\bH}\ba_{is}}{N}\right\rvert 
=\left|\frac{1}{N}\sum_{n=0}^{N-1}e^{-j2\pi{d}n\text{sin}\phi_{ir}}e^{j2\pi{d}n\text{sin}\phi_{is}}\right| 
=\frac{1}{N}\left\lvert \dfrac{\text{sin}\left({N\tau(i,i)}/{2}\right)}{\text{sin}\left({\tau(i,i)}/{2}\right)} \right\rvert \xrightarrow{\text{a.s.}} 0,
\end{align}
where $\tau(i,i)=2\pi{d}[\text{sin}\phi_{is}-\text{sin}\phi_{ir}]$, using simple results on geometric series. Almost sure convergence follows from the fact that convergence is guaranteed unless $\text{sin}\phi_{ir}=\text{sin}\phi_{is}$, an event with probability zero for continuous angular variables.
Thus, we have ${\bh_{i}^{\bH}\bh_{i}}/{N} \xrightarrow{\text{a.s.}} X_{i}$ as $N \rightarrow \infty$. Note that for random phase models, $X_i=\beta_{i}$ and traditional CH occurs where ${\bh_i^{\bH}\bh_i}/{N}\xrightarrow{\text{a.s.}} \beta_{i}$, a deterministic limit. In contrast, for complex Gaussian models, $|\gamma_{ir}|^2=\beta_{ir}|u_{ir}|^2$, which gives a random limit, as $X_{i}=\sum_{r=1}^{P}|\gamma_{ir}|^2$ is a weighted sum of exponential variables. Hence, we see that the existence of CH depends on the nature of the model for the ray coefficients. Note that the CH analysis in \cite{roy2018mimo} was for arbitrary array topologies but relied on uniform angles. The ratio of sines in (\ref{eq:signal array response vector}) appears frequently in the analysis and the more compact representation using the Dirichlet kernel is used in all subsequent derivations.
\subsection{FP (Single-Sided Massive MIMO)} \label{sec:FP With Finite Users}
For FP, results are simple following the same methodology as for CH. First, we write 
\begin{align}\label{eq:interference array response vector2}  
\dfrac{\bh_{i}^{\bH}\bh_{j}}{N}&=\frac{1}{N}\sum_{r=1}^{P}\sum_{s=1}^{P}\gamma^{*}_{ir}\gamma_{js}\ba_{ir}^{\bH}\ba_{js},
\end{align}
and then we use (\ref{eq:signal array response vector}) to show that
\begin{align}  \label{eq:interference array response vector}
\left\lvert \dfrac{\ba_{ir}^{\bH}\ba_{js}}{N}\right\rvert 
&=\frac{1}{N}\left\lvert \dfrac{\text{sin}\left({N\tau\left(i,j\right)}/{2}\right)}{\text{sin}\left({\tau\left(i,j\right)}/{2}\right)} \right\rvert =\frac{2\pi}{N}\left\lvert{D}_{\frac{N-1}{2}}\left(\tau\left(i,j\right)\right)\right\rvert\xrightarrow{\text{a.s.}} 0,
\end{align}
as $N \rightarrow \infty$, where $\tau(i,j)=2\pi{d}[\text{sin}\phi_{js}-\text{sin}\phi_{ir}]$. Hence, FP is proven very simply for all ray-based models where $\text{sin}\phi_{ir}=\text{sin}\phi_{is}$ has probability zero. A simple condition for this to hold is that the angles are continuous random variables, a property held by all proposed models.\footnote{This was demonstrated in \cite{wu2017favorable}, but only for the case of uniform distribution. In contrast, the result in (\ref{eq:interference array response vector}) is general.}  Therefore, FP, the key property enabling single-sided massive MIMO, holds for all ray distributions considered to date, such as uniform, wrapped Gaussian, and Laplacian.
\vspace{-0.03cm}
\subsection{Large System Potential (Double-Sided Massive MIMO)} \label{Large System Potential}

We analyze  LSP of ray-based channels in the limiting regime $K\rightarrow\infty$, $N\rightarrow\infty$, $N/K\rightarrow\alpha$. 
\subsubsection{Ray-based Models}

The ratio $\zeta_\textrm{LSP}$ in (\ref{SIR}) has a numerator satisfying $\mathbb{E}[{{{\lvert}{\textbf{h}}^{\text{H}}_{i}\textbf{h}_{i}\rvert}^{2}}/{N^2}]\xrightarrow{\text{a.s.}}\mathbb{E}[{X}^{2}_i]$ from Sec.~\ref{Channel Hardening} and $\mathbb{E}[X_i^2]$ is finite. Hence, LSP depends on the asymptotic properties of the denominator, $\mathbb{E}[\eta_i]$.
Here, we write,
\begin{align} \nonumber \label{eq: total interference over N square}
\mathbb{E}\left[\eta_i\right]&= \mathbb{E}\left[\frac{1}{N}\sum_{j=1, j\neq i}^{K}\frac{1}{N}\left|\sum_{r=1}^{P}\sum_{s=1}^{P}\gamma_{ir}^{*}\gamma_{js}\ba_{ir}^{\text{H}}\ba_{js}\right|^2 \right],\\
&=\frac{1}{N}\sum_{j=1, j\neq i}^{K}\sum_{r=1}^{P}\sum_{s=1}^{P}\mathbb{E}\left[|\gamma_{ir}|^2\right]\mathbb{E}\left[|\gamma_{js}|^2\right]\frac{1}{N}\mathbb{E}\left[\left|\ba_{ir}^{\text{H}}\ba_{js}\right|^2\right] 
=\beta_{i}\left(\dfrac{\sum_{j=1, j\neq i}^{K}\beta_{j}}{N}\right)\mu_{\text{ULA}},
\end{align}
using the basic properties of the $\gamma_{ir}$ terms and the notation 
\begin{align} \label{eq:mu_1} \nonumber
\mu_{\text{ULA}}\triangleq\frac{1}{N}\mathbb{E}\left[\left|\ba_{ir}^{\text{H}}\ba_{js}\right|^2\right]
&= \squeeeeze\squeeeeze
\frac{1}{N}\sum_{n=0}^{N-1}\sum_{m=0}^{N-1}
\squeeeeze\squeeeeze
\mathbb{E}
\squeeeeze
\left[e^{j2\pi{d}(m-n)\text{sin}\phi_{ir}}\right]
\squeeeeze
\mathbb{E}
\squeeeeze
\left[e^{j2\pi{d}(n-m)\text{sin}\phi_{js}}\right] \\
&=\frac{1}{N}\sum_{n=0}^{N-1}\sum_{m=0}^{N-1}\left|\mathbb{E}\left[e^{j2\pi{d}(m-n)\text{sin}\phi_{ir}}\right]\right|^2.
\end{align}
This relies on the fact that the scaling  factors are independent  and azimuth angles are i.i.d random variables. Hence, the steering vectors in $\mu_{\text{ULA}}$ are two generic but independent vectors and the subscripts $ir$ and $js$ are not strictly necessary. 
Now, we set $\hat\phi_{ir}=2\pi{d}\text{sin}\phi_{ir}$, and rewrite (\ref{eq:mu_1}) as 
\begin{align} \label{eq:mu_2} 
\mu_{\text{ULA}}&=1+2\sum_{\substack{q=1}}^{{N}-1}\left(1-\dfrac{q}{{N}}\right)|\mathbb{E}[e^{-jq\hat\phi_{ir}}]|^2.
\end{align}
In a typical drop of random user locations, the \textcolor{black}{strong} law of large numbers ensures that $\sum_{j=1}^{K}\beta_{j}/K$ \text{converges to} $\bar{\beta}$ as $K \rightarrow \infty$, where $
\bar{\beta}$ is a finite mean power. Hence, we have $\lim_{N\to\infty}\mathbb{E}\left[\eta_i\right]=(\beta_{i}\bar{\beta}/\alpha)\lim_{N\to\infty}\{\mu_{\text{ULA}}\}$. Hence, the asymptotic behavior of $\mathbb{E}\left[\eta_i\right]$ depends on $\mu_{\text{ULA}}$ which in turn depends on how quickly $|\mathbb{E}[e^{-jq\hat\phi_{ir}}]|^2$ decays. In the following theorem we present a general answer to this question.
\begin{result1} The term $\mathbb{E}[e^{-jq\hat\phi_{ir}}]$ decays as $q^{-\frac{1}{2}}$ as $q \to \infty$ with the asymptotic representation:
	\begin{align} \label{eq:finalOR3} 
	\mathbb{E}[e^{-jq\hat\phi_{ir}}] 
	\sim \dfrac{1}{\sqrt{dq}}\left(f_{\phi}\left(-\frac{\pi}{2}\right)e^{j(2\pi{d}q-\frac{\pi}{4})}+f_{\phi}\left(\frac{\pi}{2}\right)e^{-j\left(2\pi{d}q-\frac{\pi}{4}\right)}  \right), 
	\end{align}  
	where $f_{\phi}(\cdot)$ is the probability density function (PDF) of $\phi_{ir}$.
\end{result1}
\begin{proof}
	The proof is given in Appendix \ref{proofThm1}.
\end{proof}
\subsubsection{Implications of Theorem 1}
Equation (\ref{eq:finalOR3}) in Theorem 1 is a remarkable result with a simple and intuitive interpretation, wide generality  and important implications for massive MIMO: 
\begin{itemize}
	\item In terms of generality, (\ref{eq:finalOR3}) only requires the angular PDF, $f_\phi(\cdot)$,  not to have singularities which are worse than $O(x^{-1/2})$ at $x=0$. This covers all proposed models. \textcolor{black}{Certainly, all proposed models  thus far are continuous, so this condition is easily satisfied}.
	\item Interpreting (\ref{eq:finalOR3}) we see that if the end-fire direction has no energy, $f_\phi(\pm\pi/2)=0$, then $\mathbb{E}[e^{-jq\hat\phi_{ir}}]=0$.  Alternatively, if some end-fire radiation occurs then $f_\phi(\pm\pi/2) > 0$ and $\mathbb{E}[e^{-jq\hat\phi_{ir}}]=O(q^{-1/2})$.  From the above, it follows that if there is no end-fire radiation, $\mu_{\text{ULA}}$ is finite and the mean interference cannot dominate the mean power of the desired channel.
	\item Further, if there is end-fire radiation, then LSP does \textit{not} hold as $\mu_{\text{ULA}} \rightarrow \infty$. This conclusion holds by inspection of (\ref{eq:mu_2}). When $\mathbb{E}[e^{-jq\hat\phi_{ir}}]$ is $O(q^{-1/2})$, then $\sum_{q=1}^{N}|\mathbb{E}[e^{-jq\hat\phi_{ir}}]|^2$ is $O(\text{log}N)$ using well known properties of the series $\sum_{q=1}^{N}\frac{1}{q}$. Also, $\sum_{q=1}^{N}\frac{q}{N}|\mathbb{E}[e^{-jq\hat\phi_{ir}}]|^2$ is finite, so that $\mu_{\text{ULA}}$ grows to infinity, but at a very slow logarithmic rate.\footnote{Note that logarithmic growth always refers to growth that is logarithmic in $N$, the number of BS antennas.}  Note that this interference growth can be described as {\textit{critical}} as $\mathbb{E}[e^{-jq\hat\phi_{ir}}]$ decays at exactly the critical rate ($q^{-1/2}$) required for logarithmic growth. Any reduction at all in the decay rate would deliver  finite interference and therefore would enable LSP to hold.
	\item The importance of the end-fire direction can be understood as follows. For a ULA, it is not the proximity of two  ray angles that drives the interference, but the difference in the sines of the angles (see $\tau(i,j)$ in (\ref{eq:interference array response vector})). For angles close to broadside the difference in sines is largest, while near end-fire the difference is smallest, resulting in greater interference. 
\item We note that LSP does hold for i.i.d. Rayleigh fading channels. To see this we use the results in \cite{rusek2013scaling} to give $\zeta_\textrm{LSP}=\beta_i \bar{\beta}/\alpha$. Hence, ray-based models differ from classical statistical channel models in this regard. Note that the result in \cite{rusek2013scaling} is not given explicitly but can be obtained from Table $\text{I}$ in \cite{rusek2013scaling}.
	\item Overall, the result in (\ref{eq:finalOR3}) is extremely positive for double-sided massive MIMO. We have shown that in the challenging scenario where both $K$ and $N$ grow large, the interference, relative to the power of the desired channel, grows very slowly (logarithmically). Also, the scaling of this growth factor is very small, since a large amount of end-fire radiation is unlikely. Practical deployments typically employ patch elements with average look angles on the order of $+/-$ 45$^\circ$ (designed according to the downtilt angle of the array as well as the environment)\cite{tataria2019impact}. This greatly attenuates the end-fire (or near by) radiation since the patterns create explicit nulls to reject the incoming wavefronts in those directions.
	
\end{itemize}
\par
Given the power of these results, it is useful to validate the conclusions with some closed-form special cases. Note that although uniform and von-Mises (VM) angular distributions are not derived from measurements in real environments they are useful, both for validating Theorem 1 and providing exact asymptotics which are not perturbed by simulation error.

\subsubsection{Special Cases: Uniform Distribution}\label{sec:uniform}
When $\phi_{ir}\sim U[0,2\pi]$, $f_{\phi}(x) = \dfrac{1}{2\pi}$ for $-\pi \leq x \leq \pi$ and 
(\ref{eq:finalOR3}) becomes 
\begin{align} 
\mathbb{E}[e^{-jq\hat\phi_{ir}}] \label{eq:uniform}
&\sim \dfrac{1}{\pi\sqrt{dq}}\text{cos}\left(2\pi{dq}-\frac{\pi}{4}\right).
\end{align}
This limiting value is verified in the uniform case where the exact solution is known as $\mathbb{E}[e^{-jq\hat\phi_{ir}}] = \mathbb{E}[e^{-jq2\pi{d}\text{sin}\phi_{ir}}]= J_{0}(2\pi{dq})$\cite[p.~375]{abramowitz1964handbook}. For large values of $q$, $J_{0}(2\pi{dq}) \sim {\text{cos}(2\pi{dq}-\frac{\pi}{4})}/{\pi\sqrt{dq}}$\cite[p.~364]{abramowitz1964handbook}, which agrees with (\ref{eq:finalOR3}). Hence, the general asymptotic analysis in (\ref{eq:finalOR3}) is supported and the exact value of $\mathbb{E}[e^{-jq\hat\phi_{ir}}]$ can be used in (\ref{eq:mu_2}) to give the exact value of $\mu_{\text{ULA}}$.


\subsubsection{Special Cases: Von-Mises Distribution}\label{sec:VM}
The VM distribution has also been used in angular modeling \cite{wang2012performance} and has the PDF given by  
\begin{align} \label{eq:vmG}
f_{\phi}(x) = \dfrac{e^{\kappa{\text{cos}(x-\mu)}}}{2\pi{I_{0}(\kappa})}, \ -\pi \leq x \leq \pi,
\end{align}
where $\mu$ is a measure of location  and $\kappa$  is a measure of concentration. Substituting  into (\ref{eq:finalOR3}),

\begin{align} \label{eq:vmA} 
\mathbb{E}[e^{-jq\hat\phi_{ir}}] 
\sim \dfrac{e^{j2\pi{d}q}}{\sqrt{dq}}\left(\dfrac{e^{\kappa{\text{cos}(-\frac{\pi}{2}-\mu)}}}{2\pi{I_{0}(\kappa})}e^{-j\frac{\pi}{4}}+ \dfrac{e^{\kappa{\text{cos}(\frac{\pi}{2}-\mu)}}}{2\pi{I_{0}(\kappa})}e^{j\frac{\pi}{4}} \right).
\end{align} 
The exact solution can be found by integration, giving
\begin{align} \label{eq:vm} 
\mathbb{E}[e^{-jq\hat\phi_{ir}}]
&=\dfrac{I_{0}\left(\sqrt{{\kappa}^2\text{cos}^{2}(\mu)+(\kappa{\text{sin}(\mu)}-j2\pi{dq})^2}\right)}{I_{0}(\kappa)}. 
\end{align} 
Further analysis shows that (\ref{eq:vm}) is asymptotically equal to (\ref{eq:vmA}). This is briefly explained  as follows. For large $q$, the argument of $I_0(\cdot)$ in the numerator of (\ref{eq:vm}) is approximately $\kappa{\text{sin}(\mu)} - j2\pi{dq}$. Then we use the large argument approximation of $I_0(\kappa{\text{sin}(\mu)} - j2\pi{dq})$\cite[p.~364]{abramowitz1964handbook} and  simplify to give (\ref{eq:vmG}). Hence, for the VM case also, we have verified (\ref{eq:finalOR3}) and given an exact solution for $\mathbb{E}[e^{-jq\hat\phi_{ir}}]$.  

\subsection{Avoiding Interference Growth}\label{sched_ULA}
Since  the logarithmic interference growth predicted by Theorem 1  is critical, it can be removed by simple methods. For example, for any finite $K$, LSP exists and \textcolor{black}{for the asymptotic case, where $K$ is increasing,} LSP can be assured by trivial scheduling methods based on user separation. This is shown in the following.

Expanding $\eta_{i}=\sum_{j\ne i}^{K}|\bh_i^\bH\bh_j/N|^2$  in terms of the steering vectors gives
\begin{align}  \label{eq:FP infinite2}
\eta_i&=\sum_{j\ne i}\dfrac{1}{N^2}\sum_{r,s}\sum_{r^{'},s^{'}}\gamma_{ir}^{*}\gamma_{js}\gamma_{ir^{'}}\gamma^{*}_{js^{'}}\ba_{ir}^{\text{H}}\ba_{is}\ba^{\bH}_{js^{'}}\ba_{ir^{'}}.
\end{align}
From (\ref{eq:signal array response vector}), $\eta_i$ can be rewritten as
\begin{align}  \label{eq:FP infinite3}
\eta_i&=\sum_{j\ne i}\sum_{r,s,r^{'},s^{'}}\varGamma_{ijrsr^{'}s^{'}}\left(\frac{2\pi}{N}\right)^2\left\lvert{D}_{\frac{N-1}{2}}\left(\tau_1\left(i,j\right)\right)\right\rvert\left\lvert{D}_{\frac{N-1}{2}}\left(\tau_2\left(i,j\right)\right)\right\rvert,
\end{align}
where  $\sum_{r,s,r^{'},s^{'}}=\sum_{r,s}\sum_{r^{'},s^{'}}$ and $\varGamma_{ijrsr^{'}s^{'}}=\gamma_{ir}^{*}\gamma_{js}\gamma_{ir^{'}}\gamma^{*}_{js^{'}}$, $\tau_1(i,j)=2\pi{d}[\text{sin}\phi_{js}-\text{sin}\phi_{ir}]$ and   $\tau_2(i,j)=2\pi{d}[\text{sin}\phi_{js^{'}}-\text{sin}\phi_{ir^{'}}]$. Note that $\sum_{r,s,r^{'},s^{'}}\varGamma_{ijrsr^{'}s^{'}}$ is finite and for $\tau_u(i,j)\neq 0$, $\left| \dfrac{\text{sin}\left({N\tau_u(i,j)}/{2}\right)}{\text{sin}\left({\tau_u(i,j)}/{2}\right)}\right|$ is bounded for $u=1,2$. Hence, for finite $K$, $\eta_i \rightarrow 0$ for $\tau_1(i,j)\neq 0$, $\tau_2(i,j) \neq 0$ as the number of terms in the sum is equal to $K-1$ but the summand is $O(N^{-2})$. Since $P(\tau_1(i,j)=0)=P(\tau_2(i,j)=0)=0$ for continuous angular variables, it follows that $\eta_i \xrightarrow{\text{a.s.}} 0$ and LSP holds.

Next, consider the asymptotic case where the number of users ($N/K=\alpha$) is growing but users are only scheduled together if the sines of their ray angles are separated by more than a given protection level, $\epsilon>0$. Hence, $|\tau_u(i,j)| > 2\pi d \epsilon$. Using this inequality in (\ref{eq:FP infinite3}) gives
\begin{align}  \label{eq:upper}
\eta_i& \le \frac{1}{N^2{\textrm{sin}}^2(\pi d \epsilon)} \sum_{j\ne i}\sum_{r,s,r^{'},s^{'}}\varGamma_{ijrsr^{'}s^{'}}.
\end{align}
Since the right  side of (\ref{eq:upper}) converges to zero as $N \to \infty$ it follows that $\eta_i \xrightarrow{\text{a.s.}} 0$ and LSP holds.

\section{UPA: Channel Hardening, FP and LSP}
In this section, we extend the ULA results on CH, FP and LSP to a UPA. Throughout the work on LSP for a UPA we assume that $N_x \to \infty$ and $N_y \to \infty$ as $N \to \infty$. 
\subsection{Channel Hardening}
The CH results presented in (\ref{eq:FP ULA}) for a ULA remain valid for any array structure. Hence, $\bh^{\bH}_i\bh_i/N\xrightarrow{\text{a.s.}}X_i$ if $
\ba_{ir}^{\bH}\ba_{is}\xrightarrow{\text{a.s.}}0$. For the UPA we have
\begin{align} \label{eq:UPA CH} 
\dfrac{\ba_{ir}^{\bH}\ba_{is}}{N}  
=\frac{1}{N}(\ba_{irx}\otimes\ba_{iry})^{\bH}(\ba_{isx}\otimes\ba_{isy})
=\frac{1}{N}(\ba_{irx}^{\bH}\ba_{isx})(\ba_{iry}^{\bH}\ba_{isy}),
\end{align}
from basic properties of Kronecker products. Now, the cross products $\ba_{irx}^{\bH}\ba_{isx}$ and $\ba_{iry}^{\bH}\ba_{isy}$ have a similar form to the ULA. Hence, from (\ref{eq:FP ULA}) we can deduce that 
\begin{align}  \label{eq:UPA CH2}
\left\lvert \dfrac{\ba_{ir}^{\bH}\ba_{is}}{N}\right\rvert 
&=\frac{2\pi}{N_x}\left\lvert{D}_{\frac{N_x-1}{2}}\left(\tau_x\left(i,i\right)\right)\right\rvert
\frac{2\pi}{N_y}\left\lvert{D}_{\frac{N_y-1}{2}}\left(\tau_y\left(i,i\right)\right)\right\rvert,
\end{align}
where $\tau_x(i,i)=2\pi{d}_x(\text{sin}\theta_{is}\text{cos}\phi_{is}-\text{sin}\theta_{ir}\text{cos}\phi_{ir})$ and $\tau_y(i,i)=2\pi{d}_y(\text{sin}\theta_{is}\text{sin}\phi_{is}-\text{sin}\theta_{ir}\text{sin}\phi_{ir})$. As in Sec.~\ref{Channel Hardening}, $\ba_{ir}^{\bH}\ba_{is}/N$ converges to zero unless $\tau_x(i,i)=0$ or $\tau_y(i,i)=0$, an event with probability zero. Hence $\ba_{ir}^{\bH}\ba_{is}/N\xrightarrow{\text{a.s.}}0$ and $\bh_{i}^{\bH}\bh_{i}/N\xrightarrow{\text{a.s.}}X_i$.
\subsection{FP (Single-Sided Massive MIMO)}\label{FP UPA single-sided}
The ULA results in (\ref{eq:interference array response vector2}) and (\ref{eq:interference array response vector}) show that for any array, FP occurs if $\dfrac{\ba_{ir}^{\bH}\ba_{js}}{N}\xrightarrow{\text{a.s.}}0$ as $N\rightarrow\infty$. Following the same calculation as in (\ref{eq:UPA CH2}) gives
\begin{align}  \label{eq:UPA FP single-sided}
\left\lvert \dfrac{\ba_{ir}^{\bH}\ba_{js}}{N}\right\rvert 
&=\frac{2\pi}{N_x}\left\lvert{D}_{\frac{N_x-1}{2}}\left(\tau_x\left(i,j\right)\right)\right\rvert
\frac{2\pi}{N_y}\left\lvert{D}_{\frac{N_y-1}{2}}\left(\tau_y\left(i,j\right)\right)\right\rvert,
\end{align}
where $\tau_x(i,j)=2\pi{d}_x(\text{sin}\theta_{js}\text{cos}\phi_{js}-\text{sin}\theta_{ir}\text{cos}\phi_{ir})$ and $\tau_y(i,j)=2\pi{d}_y(\text{sin}\theta_{js}\text{sin}\phi_{js}-\text{sin}\theta_{ir}\text{sin}\phi_{ir})$.  Since $P(\tau_x(i,j)=0)=P(\tau_y(i,j)=0)=0$ for continuous angular variables, it follows that $\eta_i\xrightarrow{\text{a.s.}} 0$ and FP holds.

\subsection{Large System Potential (Double-Sided Massive MIMO)} \label{log UPA}
The existence of LSP for a UPA depends on whether the expression  $\mu_{\text{UPA}}=\frac{1}{N}\mathbb{E}\left[|\ba_{ir}^{\text{H}}\ba_{js}|^2\right]$ converges or not. This is analyzed in {Theorem 2}.
\begin{result2} The term $\mu_{\text{UPA}}$ for a UPA antenna structure grows logarithmically with the following representation:
\begin{align} \label{eq:nu_{v,w}} 
\mu_{\text{UPA}}&=\sum_{v=1-N_x}^{N_x-1}\sum_{w=1-N_y}^{N_y-1}\left(1-\frac{|v|}{N_x}\right)\left(1-\frac{|w|}{N_y}\right)|M_{v,w}|^2,
\end{align}  
where $M_{v,w}=\mathbb{E}[e^{-j2\pi\nu_{v,w}\text{sin}(\theta)\text{cos}(\phi-\Delta_{v,w})}]$, $\nu_{v,w}=\sqrt{v^2d^2_x+w^2d^2_y}$, and $\Delta_{v,w}=\text{tan}^{-1}\left(\dfrac{wd_y}{vd_x}\right)$. 
\end{result2}
\begin{proof}
	The proof is given in Appendix \ref{proofThm2}.
\end{proof}
\subsubsection{Implications of Theorem 2}
As in the ULA case,  App.~\ref{proofThm2} shows that the interference growth is critical and any reduction in the rate of interference accumulation will lead to LSP holding. The result holds for any continuous angular distributions so is extremely general. Note that a ULA would normally aim to null the end-fire direction (see (\ref{eq:finalOR3})) and perfect nulling would avoid the interference growth. In contrast, a similar argument using (\ref{UPA:M3}) shows that $f_{\theta}(x)$ must equal zero for $x$ in $\{-\pi/2, 0, \pi/2\}$ in order to avoid interference growth. For a vertical UPA, to null the broadside direction  is clearly unsuitable as it requires nulling the dominant azimuth plane ($\theta=0$). In general for all types of UPA, a more symmetric structure means that there are no sets of special angles which avoid interference growth and for which the radiation is unwanted.
Since the proof is complex, it is instructive to look at the uniform case where a closed-form result for $\mu_{\text{UPA}}$ can be derived and shown to grow logarithmically as in {Theorem 2}. This is shown in the following.
\subsubsection{Special Case: Uniform Distribution}
We derive a closed-form equation for $\mu_{\text{UPA}}$ in the simplest uniform case, where the
azimuth angle is $\phi_{ir}\sim U[0,2\pi]$ and the elevation angle is $\theta_{ir}\sim U[-\frac{\pi}{2},\frac{\pi}{2}]$. In this scenario, the following result applies.
\begin{result3} 
The value  of $\mu_\text{UPA}$ for angles uniform in azimuth and in elevation is given by
	\begin{align}\label{eq:upa uniform}
	\mu_\text{UPA}&=\frac{1}{N}\sum_{r_x=1}^{N_x}\sum_{r_y=1}^{N_y}\sum_{s_x=1}^{N_x}\sum_{s_y=1}^{N_y} 
	J_0^4\left(\pi\sqrt{d_y^2(r_y-s_y)^2+d_x^2(r_x-s_x)^2}\right).
	\end{align}
\end{result3}
\noindent where the proof is given in Appendix \ref{mu_for_UPA}. In Appendix \ref{The logarithm growth of Uniform Distribution for UPA}, this result is used to demonstrate the logarithmic growth of $\mu_{\text{UPA}}$.

\subsubsection{Avoiding Interference Growth}
As for the ULA, interference growth can be avoided by using finite $K$ or simple scheduling. This is shown in the following. For a UPA, substituting the associated steering vectors into (\ref{eq:FP infinite2}) gives 
\begin{align}  \label{eq:UPA FP double-sided} 
\eta_i&=\sum_{j\ne i}\dfrac{1}{N^2}\sum_{r,s,r^{'},s^{'}}\varGamma_{ijrsr^{'}s^{'}}\ba_{irx}^{\text{H}}\ba_{isx}\ba_{iry}^{\text{H}}\ba_{isy}\ba^{\bH}_{js^{'}x}\ba_{ir^{'}x}\ba^{\bH}_{js^{'}y}\ba_{ir^{'}y}.
\end{align}
As in (\ref{eq:UPA CH2}), all of the four cross products of steering vectors in $x$ and $y$ domains have representations as ratios of sine functions. Hence,
\small
\begin{align}  \label{eq:UPA FP infinite} \nonumber
\eta_i&=\sum_{j\ne i}\sum_{r,s,r^{'},s^{'}}\frac{16\pi^4}{N^2}\varGamma_{ijrsr^{'}s^{'}}\left\lvert{D}_{\frac{N_x-1}{2}}\left(\tau_x\left(i,j\right)\right)\right\rvert\left\lvert{D}_{\frac{N_y-1}{2}}\left(\tau_y\left(i,j\right)\right)\right\rvert\left\lvert{D}_{\frac{N_x-1}{2}}\left(\tau_x^{'}\left(i,i\right)\right)\right\rvert\left\lvert{D}_{\frac{N_y-1}{2}}\left(\tau_y^{'}\left(j,j\right)\right)\right\rvert,
\end{align}
\normalsize
where $\tau_x(i,j)$ and $\tau_y(i,j)$ are defined in Sec.~\ref{FP UPA single-sided}, $\tau_x^{'}(i,i)=2\pi{d}_x(\text{sin}\theta_{is^{'}}\text{cos}\phi_{is^{'}}-\text{sin}\theta_{ir^{'}}\text{cos}\phi_{ir^{'}})$ and $\tau_y^{'}(j,j)=2\pi{d}_y(\text{sin}\theta_{js^{'}}\text{sin}\phi_{js^{'}}-\text{sin}\theta_{jr^{'}}\text{sin}\phi_{jr^{'}})$. As before, for finite $K$ and $N$ growing,  $\eta_i \to 0$ unless at least one of $\tau_x(i,j)$, $\tau_y(i,j)$, $\tau_x'(i,i)$, $\tau_y'(j,j)$ is zero, an event with probability $0$. Hence,  $\eta_i \xrightarrow{\text{a.s.}} 0$ as $N\rightarrow\infty$ so that LSP holds. In the asymptotic case where $K\rightarrow\infty$, considering scheduling using a similar protection threshold as used in Sec.~\ref{sched_ULA} where users are only selected if $\textrm{min}(\tau_x(i,j),\tau_y(i,j),\tau_x^{'}(i,i),\tau_y^{'}(j,j))>2\pi d \epsilon$. With this approach, we see that
\begin{equation}\label{prot_ura}
\eta_i < \sum_{j\ne i}^{K}\frac{1}{N^2\textrm{sin}^4(\pi d \epsilon)}\sum_{r,s,r^{'},s^{'}}\varGamma_{ijrsr^{'}s^{'}}.
\end{equation}
Since the right side of (\ref{prot_ura}) converges to zero as $N \to \infty$ ($N/K=\alpha$),  $\eta_i \xrightarrow{\text{a.s.}} 0$ and LSP holds.

\section{Numerical Results}\label{Numerical Results}
\begin{figure}[ht]

\centering\myincludegraphics{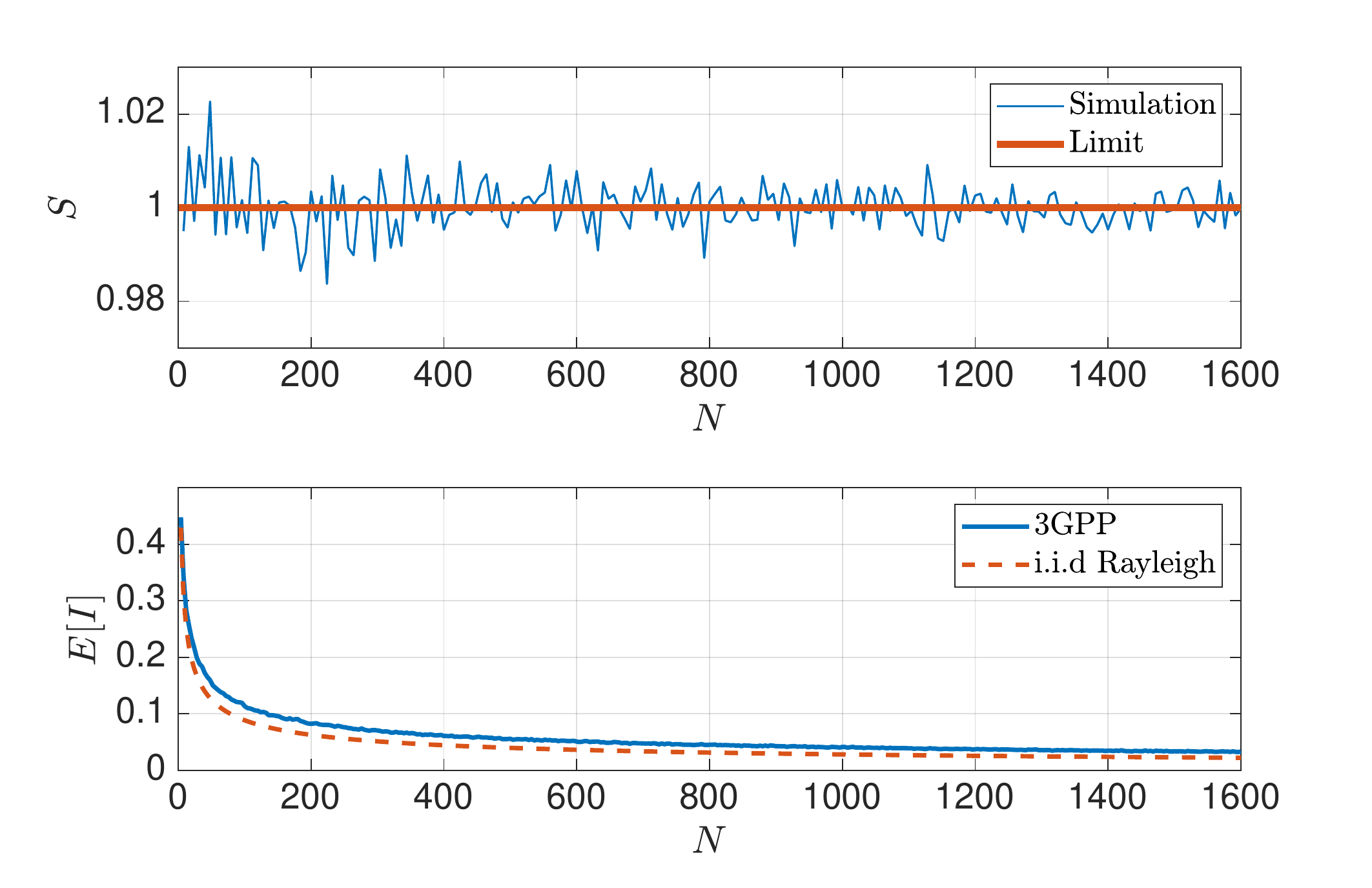}
	\caption{Channel hardening and FP (3GPP angular parameters).}
	\label{SignalInter1600}	
\end{figure}

\begin{figure}
\centering\myincludegraphics{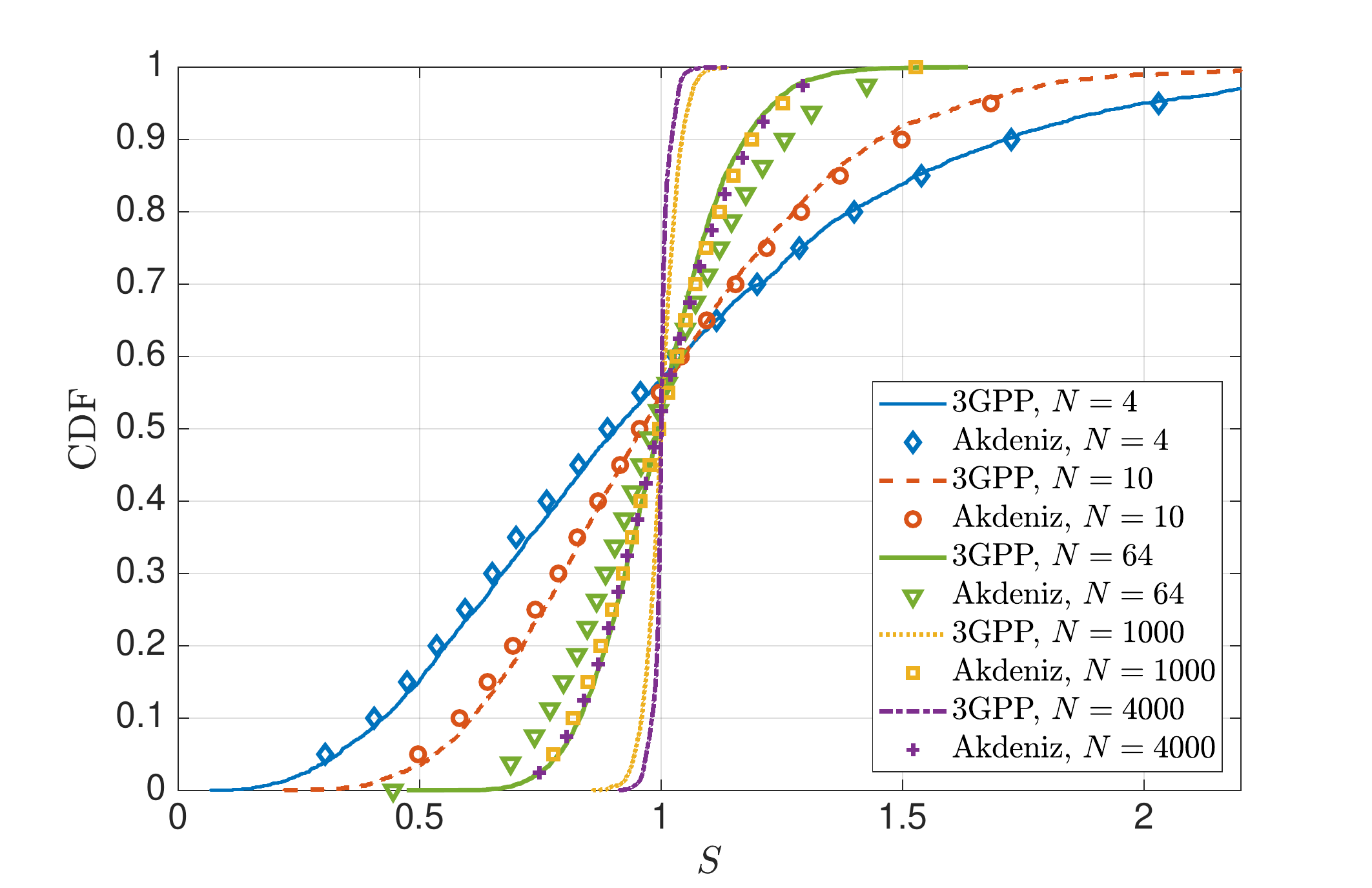}
	\caption{Channel hardening for two types of channel models.}
	\label{signalpowerComparisonOriginal0411}
\end{figure}

In Fig.~\ref{SignalInter1600} we demonstrate the CH and FP results for the ULA discussed in Sec.~\ref{sec: fixedSV} for $K=2$ and an increasing number of antennas. We adopt the non-line-of-sight (NLOS) 3GPP angular and cluster parameters in \cite{ref26}. The number of clusters is $C=20$, and the number of subpaths per cluster is $L=20$. Referring to the channel model in (\ref{SVchannel}), $P=CL$. Each subpath angle of arrival (AoA) is modeled by a  central cluster angle with a Gaussian distribution (zero mean and a standard deviation of $76.5^\circ$) plus a subray offset angle which is Laplacian with a standard deviation of $15^\circ$. 
We assume $\beta_{1}=\beta_{2}=1$ and subrays with equal powers.\footnote{Equal ray powers are adopted for simplicity in  Fig.~\ref{SignalInter1600} and  Fig.~\ref{signalpowerComparisonOriginal0411} for initial verification of the FP and CH results, $\beta_{ir}={1}/{{CL}}$, and phases are uniformly distributed, $\phi_{ir}\sim U[0, 2\pi]$.} From the upper plot of Fig.~\ref{SignalInter1600}, we  see that the normalized power in the desired channel, $S=|{\bh_{i}^{\bH}\bh_{i}}|/{N} \approx 1$ for large numbers of antennas. Similarly, the lower plot shows the mean of  interference term, $I=|\bh_{i}^{\bH}\bh_{j}|/N$ decreasing to zero as $N \rightarrow \infty$. Note that Fig.~\ref{SignalInter1600} plots simulations of $\mathbb{E}[I]$ for the ray-based model and analytical values of $\mathbb{E}[I]$ for i.i.d. Rayleigh fading (where each element of $\textbf{H}$ is an independent  $\mathcal{CN}(0,1)$ random variable) so that the variations do not obscure the trend. As expected, the convergence to FP is slower for the ray-based model but the initial rate of convergence is similar for both channels.
 Hence, both CH and FP  are shown to occur for a typical parameter set as predicted by the analysis. Fig.~\ref{SignalInter1600} shows CH and FP occurring for a clustered channel model with wrapped Gaussian central cluster angles and Laplacian offsets. This numerical example is useful as it verifies the analysis for a commonly used ray-based model structure. The analysis goes much further and proves the existence/non-existence of CH and the existence of FP for all ray-based of the form in (\ref{SVchannel}) for a comprehensive range of ray distributions. These observations are in line with the CH measurements reported in \cite{gunnarsson2018channel}.

Fig.~\ref{signalpowerComparisonOriginal0411} shows the power of the desired channel will either converge to a constant or a random variable, verifying the analysis in Sec.~\ref{Channel Hardening}. We assume the same model as in Fig.~\ref{SignalInter1600} but with two possibilities for the ray coefficients, $\gamma_{ir}$. The Akdeniz model \cite{ref53} uses a complex Gaussian variable for $
\gamma_{ir}$, while the 3GPP model \cite{ref26} uses a random phase. As shown in Fig.~\ref{signalpowerComparisonOriginal0411}, as the number of antennas grows, the cumulative distribution function (CDF) of the normalized desired channel power, $S={\bh_i^{\bH}\bh_i}/{N}$, with the Akdeniz model remains almost the same, indicating convergence to a random variable. In contrast, with the random phase model of 3GPP the CDF converges to a step function indicating that $S$ converges to a constant. Hence, as shown in Sec.~\ref{Channel Hardening}, CH can occur for ray-based channels, depending on the ray coefficient models. 
	
%

\par 
\begin{figure}[ht]
\centering\myincludegraphics{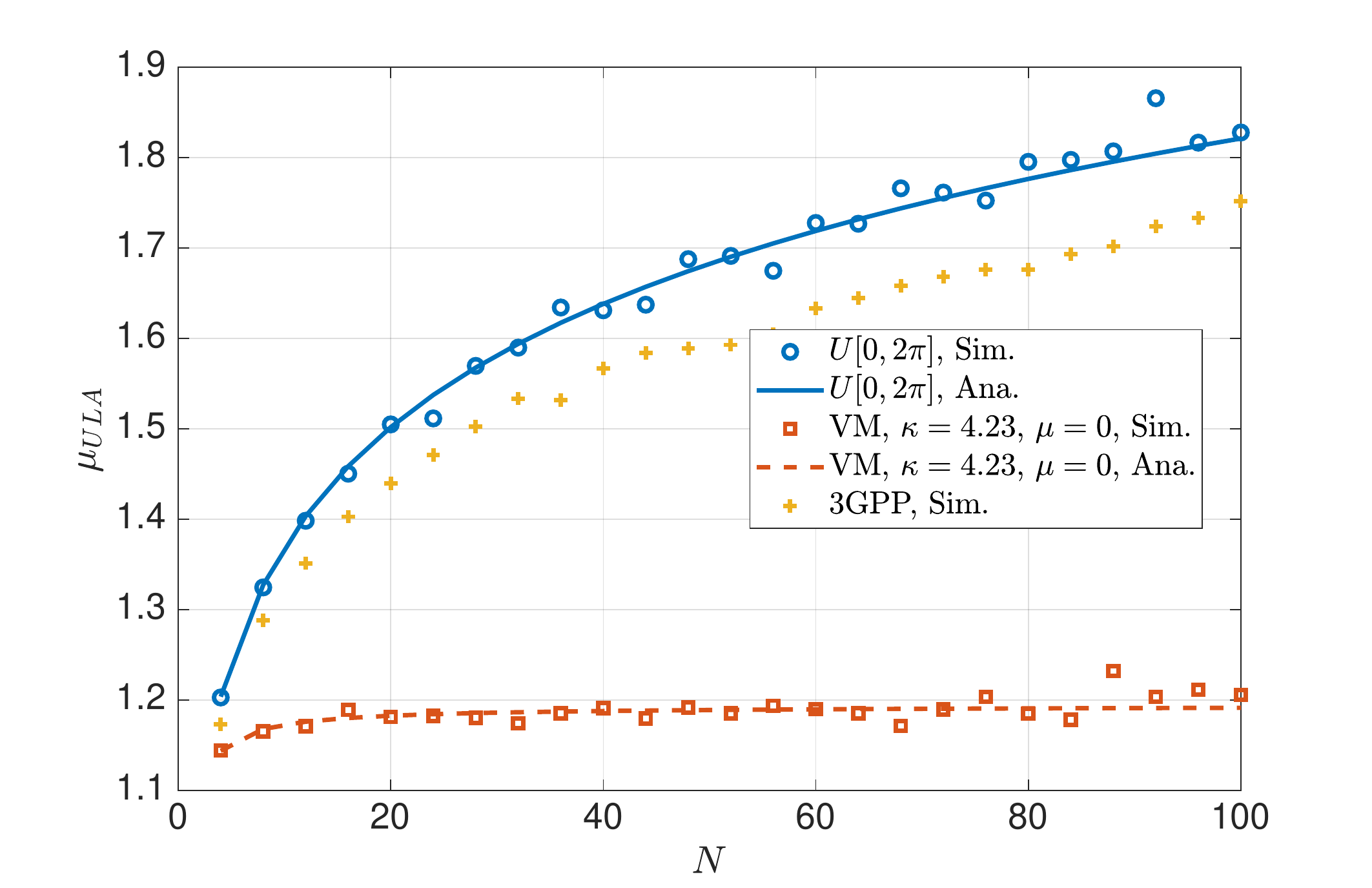}
	\caption{$\mu_\text{ULA}$ vs $N$ for three different angular distributions.}
	\label{muNDall}	
\end{figure}
In Fig.~\ref{muNDall}, we show both the simulated and analytical results for $\mu_\text{ULA}$ with uniform and VM distributions using the results in Secs.~\ref{sec:uniform} and \ref{sec:VM}. We also show simulated values of $\mu_\text{ULA}$ by adopting the angular parameters of the 3GPP model in \cite{ref26} as in Fig.~\ref{SignalInter1600}. The number of antennas and users are growing at the same ratio ${N}/{K}=\alpha=2$, while 
$\phi_{ir} \sim U[0,2\pi]$ for the uniform model and $\kappa=4.23$  (for $30^\circ$ angle spread) and $\mu=0$ for VM. From Fig.~\ref{muNDall} we see that the analysis agrees well with simulation for both uniform and VM models. We also note that the growth rates of $\mu_\text{ULA}$ are different for all three models, due to the differences in the AoA distributions. Next, we give more details of the growth rate with regard to angular distributions.

\begin{figure}[ht]
	\centering\myincludegraphics{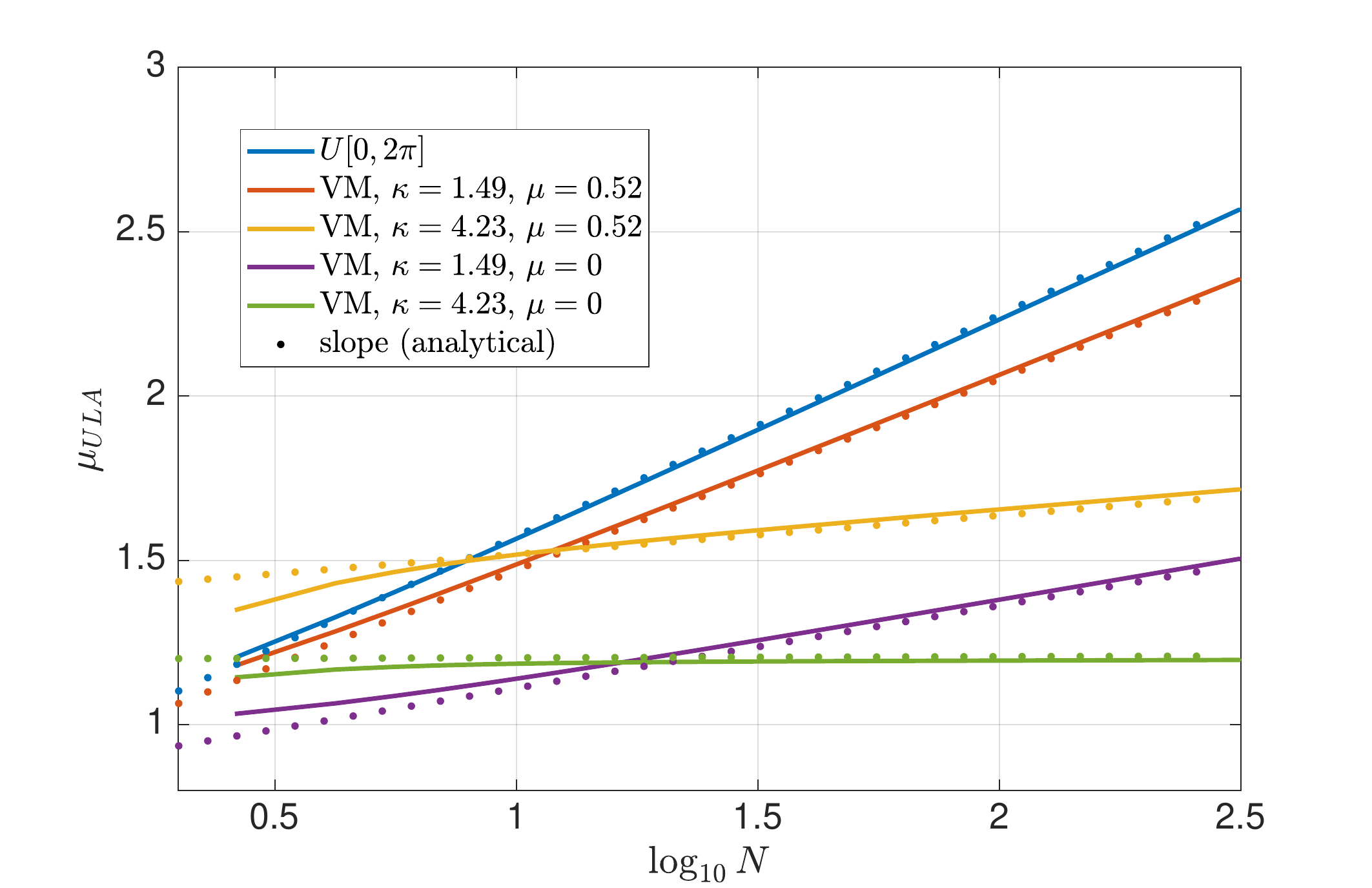}
	\caption{Logarithmic growth of $\mu_\text{ULA}$.}
	\label{muNDallComparisonVMuniform}	
\end{figure}

In Fig.~\ref{muNDallComparisonVMuniform}, we demonstrate the logarithmic growth rate of $\mu_\text{ULA}$ against the number of antennas, $N$ \textcolor{black}{($\alpha=2$)}, for VM and uniform models with  parameters as shown in the legend. The analytical results in Secs.~\ref{sec:uniform} and \ref{sec:VM} were used in generating $\mu_\text{ULA}$ for the uniform and VM distributions, respectively. Although the analysis in Theorem 1 predicted logarithmic growth for $\mu_{\text{ULA}}$, this is hard to verify from Fig.~\ref{muNDall}. Hence, we substitute (\ref{eq:finalOR3}) into (\ref{eq:mu_2}) and identify the dominant component of $\mu_\text{ULA}$ giving $\mu_{\text{ULA}}\sim m_\text{slope}\text{log}(N)+C_0$, where $C_0$ is a constant and 
\begin{align} 
{m}_\text{slope}=\dfrac{2(f_{{\phi}}^2(\frac{\pi}{2})+f_{{\phi}}^2(-\frac{\pi}{2}))}{d}.
\end{align}
Hence, ${m}_\text{slope}$ determines how quickly $\mu_\text{ULA}$ will grow. The uniform distribution has the highest interference growth rate, which is ${m}^{\text{uniform}}_\text{slope}={(\pi^2d)}^{-1}$. For the VM model, the slope depends on $\kappa$ and $\mu$. In Fig.~\ref{muNDallComparisonVMuniform} we observe that $\mu_\text{ULA}$ is clearly logarithmic in $N$, as predicted, and that the slope is correctly identified by (\ref{eq:finalOR3}),  as shown by the dotted lines which have slope $m_\text{slope}$. 
\par 
As well as verifying the logarithmic growth, Fig.~\ref{muNDallComparisonVMuniform} demonstrates some interesting angular properties. For both $\kappa=4.23$ ($\text{angle spread}=30^\circ$) and $\kappa=1.49$ ($\text{angle spread}=60^\circ$), $\mu_\text{ULA}$ decreases as $\mu$ is reduced from $\mu=0.52$ ($30^\circ$) to $\mu=0$. This is because shifting the mean towards broadside reduces the interference inflation that occurs near end-fire. Secondly, for both $\mu=0$ and $\mu=0.52$ there is a cross-over as $N$ increases. For small $N$, increased angular spread is beneficial as it spreads the rays and reduces the chance of high interference caused by rays in close proximity. However, for high $N$ the higher angular spread puts more probability near end-fire and this begins to dominate and causes higher interference.


\begin{figure}[ht]
\centering\myincludegraphics{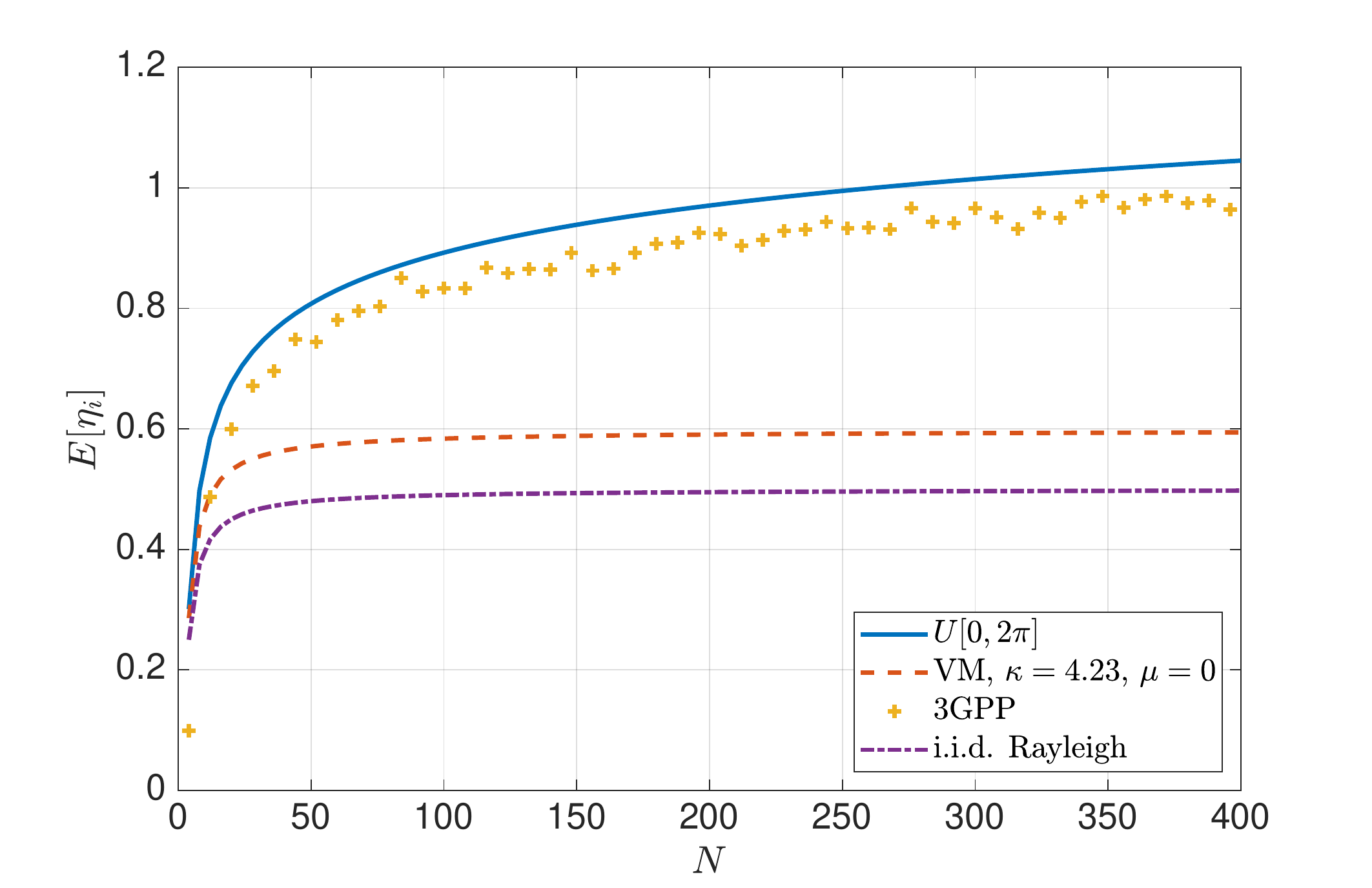}
	\caption{$\mathbb{E}[\eta_{i}]$ vs $N$ for three angular distributions for ULA and Rayleigh fading.}
	\label{allTotalULA}	
\end{figure}

In Fig.~\ref{allTotalULA}, we confirm via simulation for the 3GPP parameters and via analysis for the uniform and VM models that the mean global interference term, $\mathbb{E}[\eta_i]$, grows logarithmically as predicted by the analysis in Sec.~\ref{Large System Potential}. For the uniform case, $\phi_{ir}\sim U[0,2\pi]$, for VM, $\kappa=4.23$, $\mu=0$, and for 3GPP we use the parameters considered in Fig.~\ref{SignalInter1600}. For the uniform and VM models all user link gains and ray powers are equal, $\beta_{ir}=({CL})^{-1}$. For the 3GPP parameters, we also consider unequal ray powers and unequal user link gains. To avoid the substantial extra variation caused by shadowing models with large arrays we employ a simple deterministic model for these powers. The link gains decay exponentially from user $1$ to user $K$ such that $\beta_{K}=\frac{1}{10}\beta_{1}$ and the cluster powers behave similarly. The desired user is then randomly allocated one of the $K$ distinct link gains \textcolor{black}{($\alpha=2$)}. The levels are then adjusted to give the same total power as in the uniform and VM models and subrays in a particular cluster all have the same power as assumed in \cite{ref3GPP}. Fig.~\ref{allTotalULA} shows the same logarithmic growth as Fig.~\ref{muNDall}, confirming the analysis. The growth of the VM curve is hard to see on this scale but is clear in Fig.~\ref{muNDallComparisonVMuniform}. Also shown in Fig.~\ref{allTotalULA} are results for the i.i.d. Rayleigh case where all users have unit link gain. As discussed in Sec.~\ref{sec:FP With Finite Users} and \cite{rusek2013scaling}, for Rayleigh fading $\eta_{i} \rightarrow 1/\alpha$.  Since $\alpha=2$ in Fig.~\ref{allTotalULA}  we observe convergence to $0.5$. In comparison, it is clear that the uniform and 3GPP models continue to grow. Hence, the opposite behavior occurs  with ray-based models compared to Rayleigh. The growth of the VM curve is hard to see on this scale but is clear in Fig.~\ref{muNDallComparisonVMuniform}.

\begin{figure}[ht]
	\centering\myincludegraphics{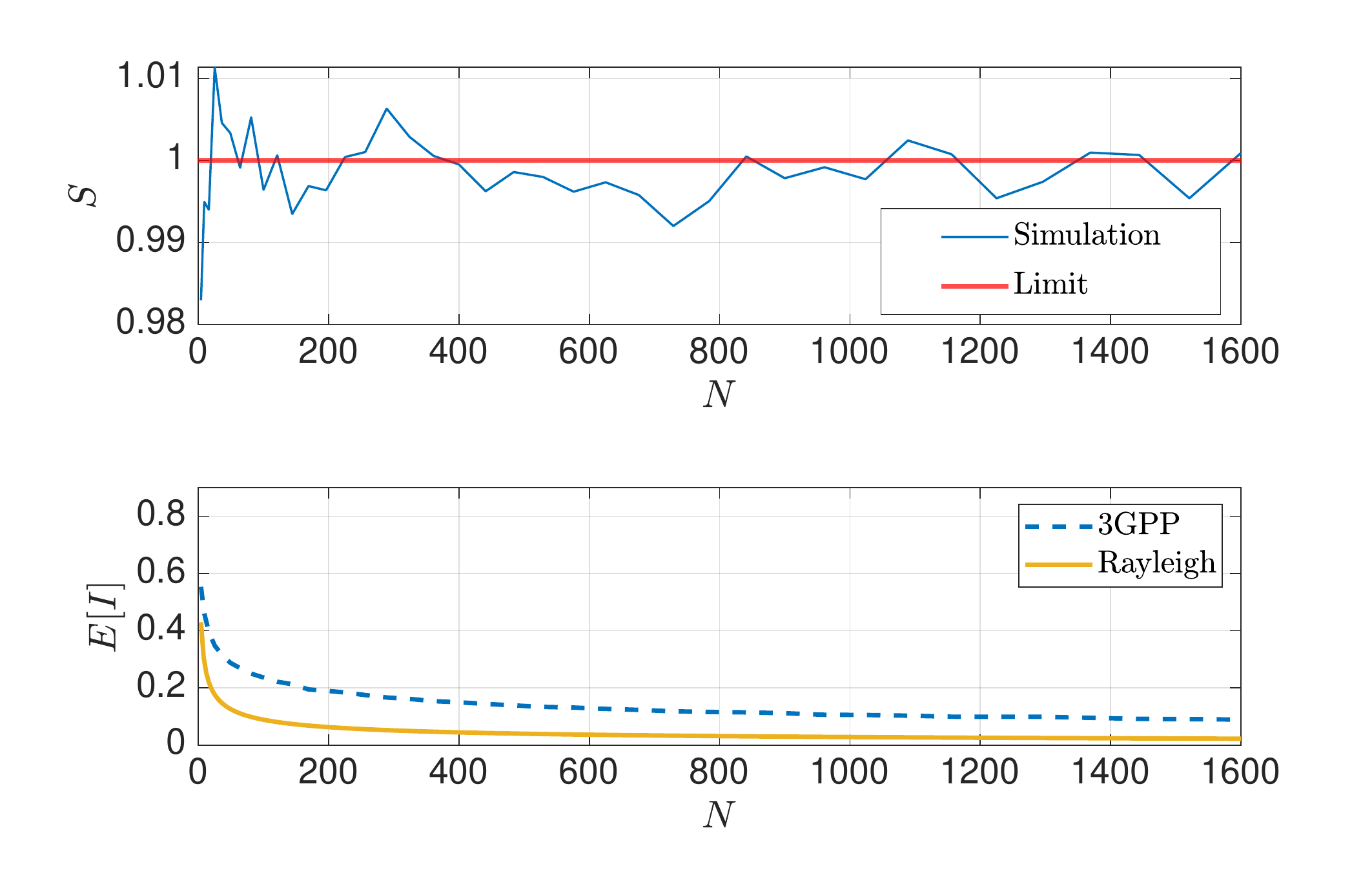}
	\caption{Channel hardening and FP for UPA.}
	\label{UPAfigureOneFPSignal}	
\end{figure}

In Fig.~\ref{UPAfigureOneFPSignal}, we evaluate the CH and FP results for a UPA with $K=2$ and an increasing number of antennas. We adopt the angular models from \cite{sangodoyin2018cluster}. The azimuth central angle follows a wrapped Gaussian distribution ($\phi_{c}\sim\mathcal{N}(0^{\circ},\sigma_c^{2})$) with a Laplacian offset distribution ($\Delta_{c,s}\sim\mathcal{L}(1/\sigma_s)$). Both the central and offset angles for elevation have Laplacian distributions with $\theta_{c}\sim\mathcal{L}(1/\hat\sigma_c)$ (centred on $90^{\circ}$) and $\delta_{c, s}\sim\mathcal{L}(1/\hat\sigma_s)$, respectively. We adopt the following parameters: the number of clusters is $C=20$, the number of subrays within a cluster is $L=20$, $\sigma_c=31.64^{\circ}$ and  $\sigma_s=24.25^{\circ}$ from \cite{ref3GPP} and we use the upper $90\%$ of the lognormally distributed values of $\hat\sigma_c$  and $\hat\sigma_s$ from \cite{sangodoyin2018cluster} in order to give a wide angular spread, which is $6.12^{\circ}$ and $1.84^{\circ}$, respectively (Scenario Wide). As in Fig.~\ref{SignalInter1600} and Fig.~\ref{signalpowerComparisonOriginal0411}, equal subray powers are assumed and $\beta_{ir}={1}/{(CL)}$ for simplicity. From Fig.~\ref{UPAfigureOneFPSignal}, we see that, similar to ULA, FP and CH also occur for the UPA structure. However, the gap between the 3GPP and Rayleigh channel in Fig.~\ref{UPAfigureOneFPSignal} is wider than for a ULA. Hence, the smaller azimuth footprint of the UPA slows down FP.
  
\begin{figure}[ht]
	\centering\myincludegraphics{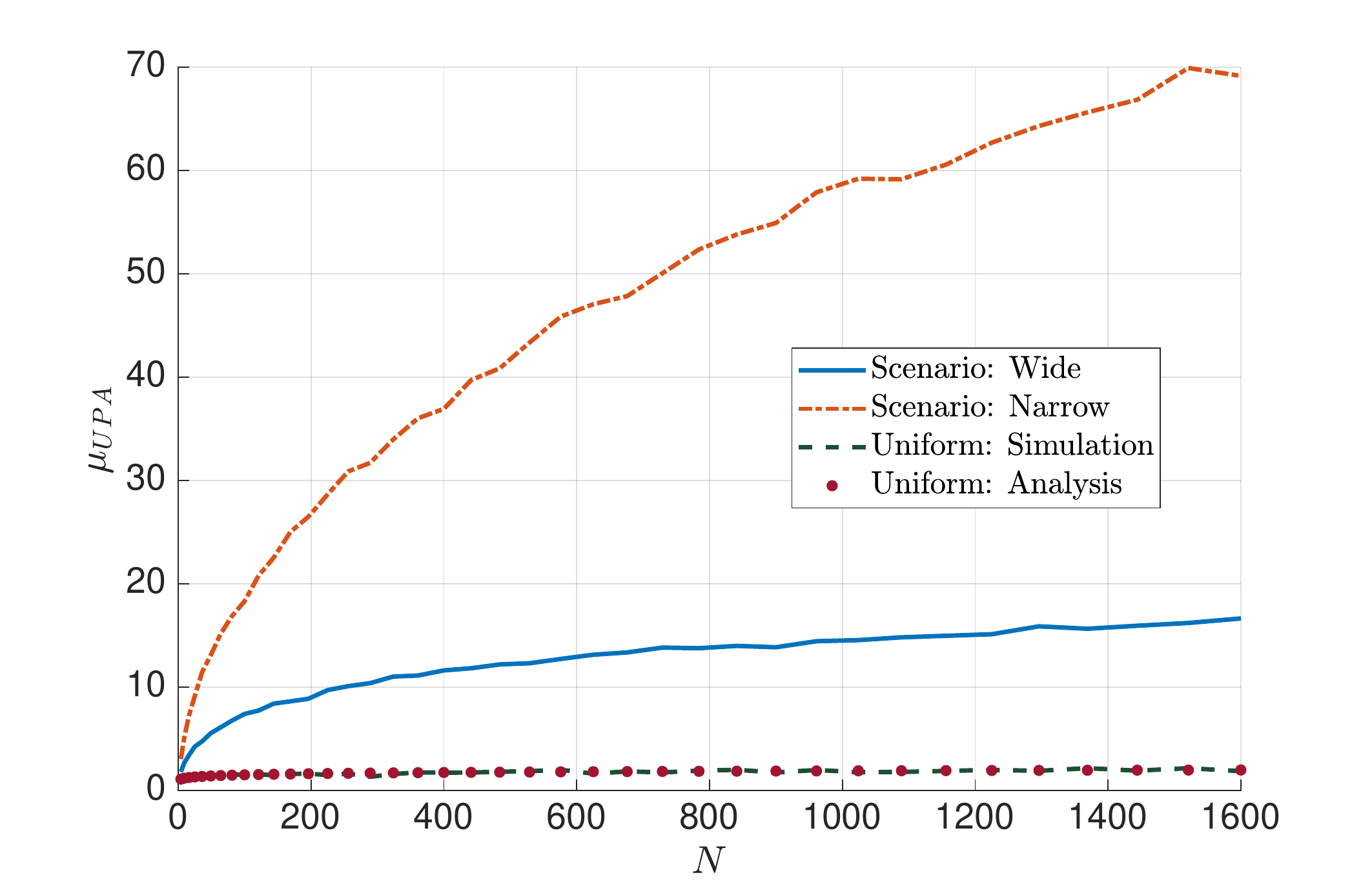}
	\caption{$\mu_\text{UPA}$ vs $N$ for wide and narrow angular distributions for UPA.}
	\label{UPAfigure2mu}	
\end{figure}

 Fig.~\ref{UPAfigure2mu} shows the simulated and analytical results for $\mu_\text{UPA}$ with a uniform angular distribution for both azimuth ($U[0, 2\pi]$) and elevation angles ($U[0, \pi]$). The analytical results are from Sec.~\ref{log UPA}. We see that the analysis agrees well with the simulation. We also show the  simulated values of $\mu_\text{UPA}$ for two scenarios. Scenario Wide  uses the angular parameters in Fig.~\ref{UPAfigureOneFPSignal}. Scenario Narrow uses $C=3$,  $L=16$, $\sigma_c=14.4^{\circ}$ and  $\sigma_s=6.24^{\circ}$ and the lower $10\%$ of the lognormally distributed values of $\hat\sigma_c$  and $\hat\sigma_s$ from \cite{sangodoyin2018cluster}. As with the ULA, the growth rates of $\mu_\text{UPA}$ are different for all three models, due to the differences in  azimuth and elevation angular distributions. 
\begin{figure}[ht]
	\centering\myincludegraphics{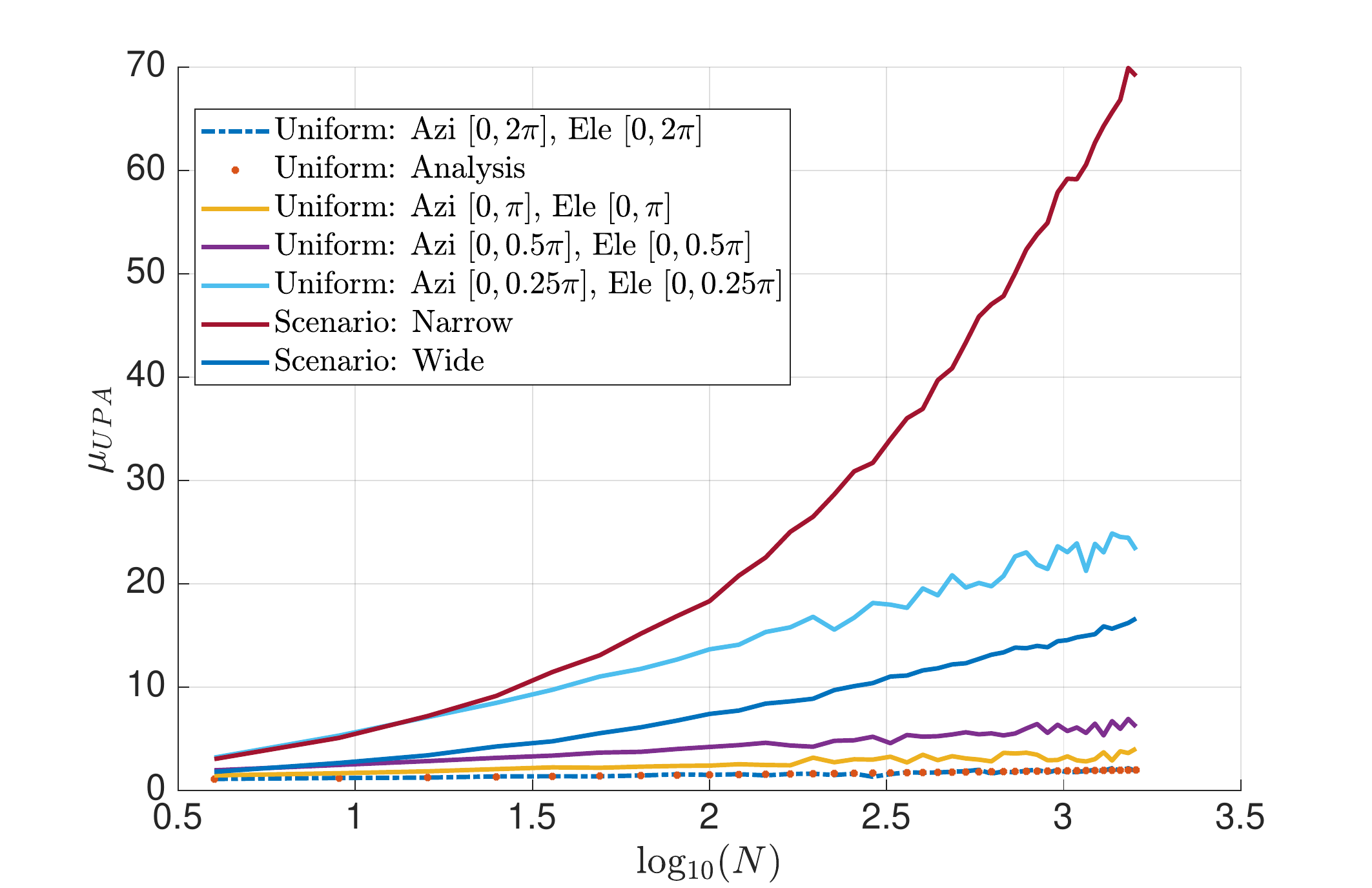}
	\caption{Logarithmic growth of $\mu_\text{UPA}$ for UPA.}
	\label{figure3logGrowthUPA}	
\end{figure}

Fig.~\ref{figure3logGrowthUPA} shows the logarithmic growth of $\mu_\text{UPA}$ with different angular spreads in both azimuth and elevation angles. As we can see, the narrower the angular spread, the quicker $\mu_\text{UPA}$ grows. Note that the lower curves have stabilized and show linear growth with $\text{log}_{10}(N)$ while the narrow scenario has not yet reached the high $N$ regime where  logarithmic growth observed. 
\begin{figure}[ht]
	\centering\myincludegraphics{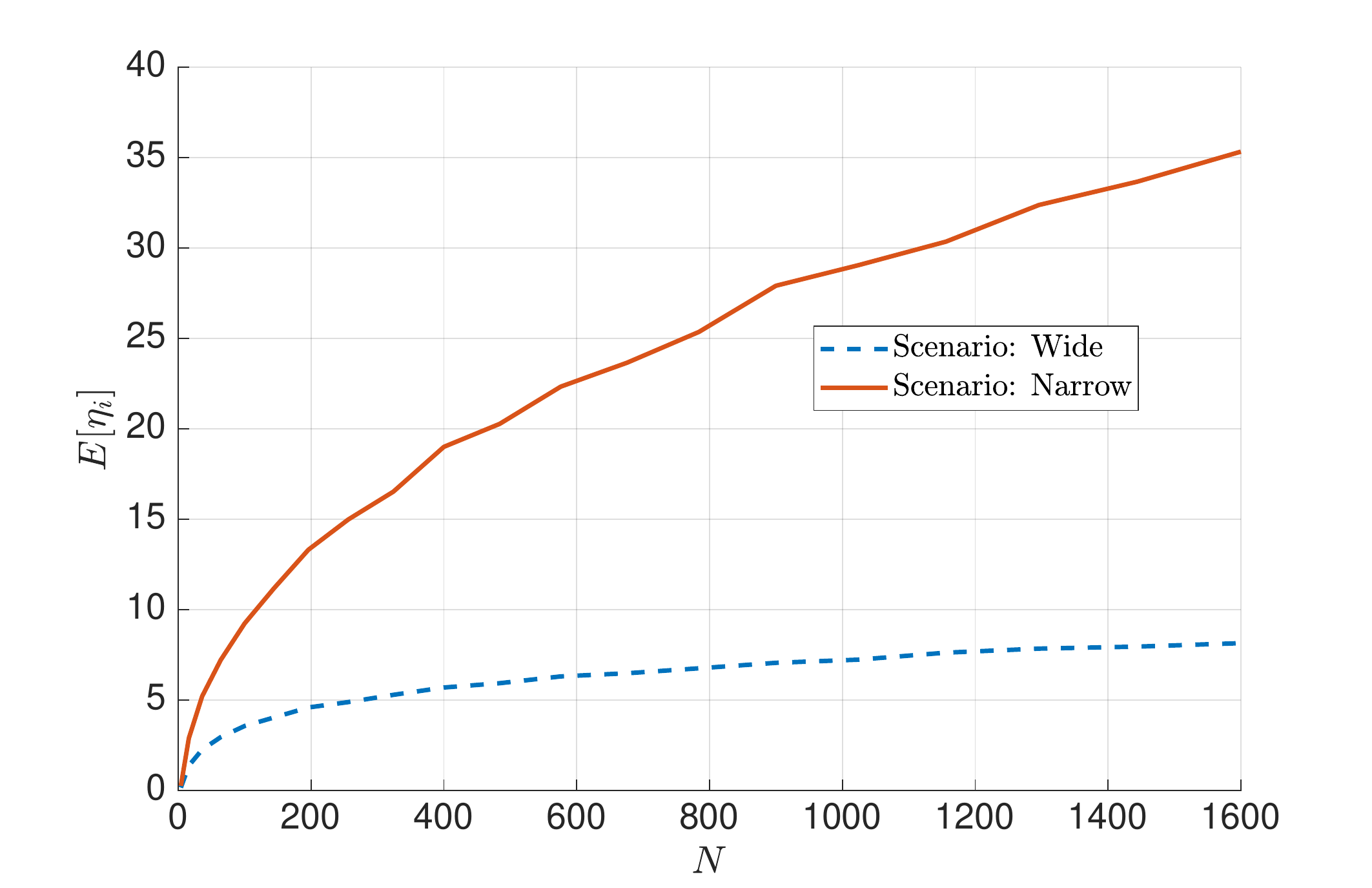}
	\caption{$\mathbb{E}[\eta_{i}]$ vs $N$ for two angular distributions for UPA.}
	\label{expBeta}	
\end{figure}

\begin{figure}[ht]
	\centering\myincludegraphics{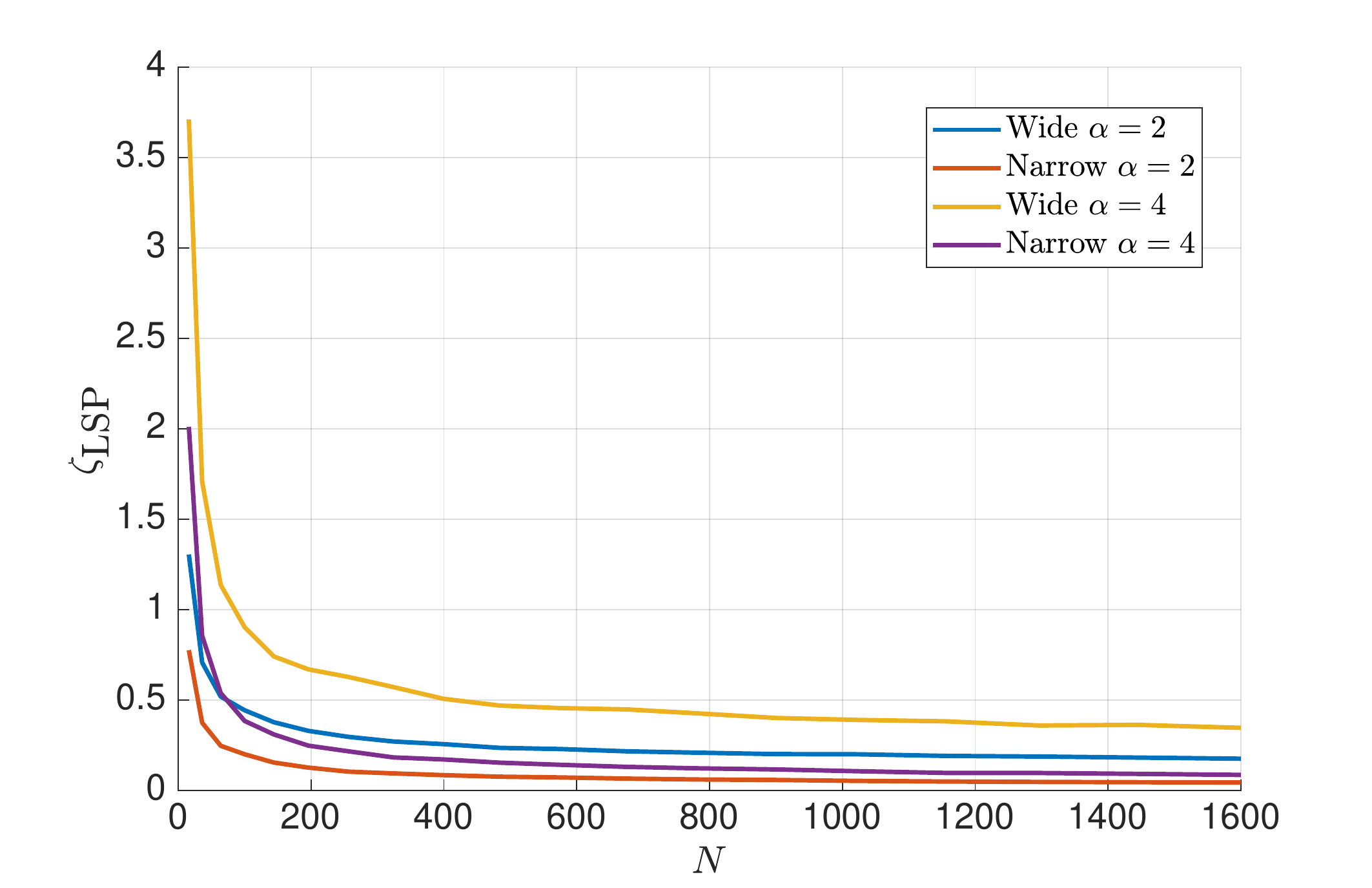}
	\caption{$\zeta_\textrm{LSP}$ vs $N$ for two angular distributions for UPA.}
	\label{LSP}	
\end{figure}

In Fig.~\ref{expBeta}, we confirm via simulation for the 3GPP parameters (adopting Scenario Wide and Scenario Narrow parameters of Fig.~\ref{UPAfigure2mu}) that the mean global interference, $\mathbb{E}[\eta_i]$, grows logarithmically as predicted by the analysis in Sec.~\ref{log UPA}. We consider unequal ray powers and unequal user link gains, as in Fig.~\ref{allTotalULA}.
Figs.~\ref{UPAfigure2mu}-~\ref{expBeta} show the logarithmic growth of the interference which is sufficient to show that  $\zeta_{\textrm{LSP}} \to 0$ as $N \to \infty$. For completeness, in Fig.~\ref{LSP} we also plot 
$\zeta_{\textrm{LSP}}$ against $N$ for the same parameters used in Fig.~\ref{expBeta} with the exception that both $\alpha=2$ and $\alpha=4$ are shown. On the $x$-axis, the array size, $N$, ranges from $16$ (a $4 \times 4$ horizontal UPA) to $1600$ (a $40 \times 40$ array). As expected, $\zeta_{\textrm{LSP}}$ decays more quickly for
the more challenging scenarios, ie. smaller $\alpha$ and narrower angular spread. After the initial drop the decay to zero is slow as the interference growth is only logarithmic.\\
\begin{figure}[ht]
	\centering\myincludegraphics{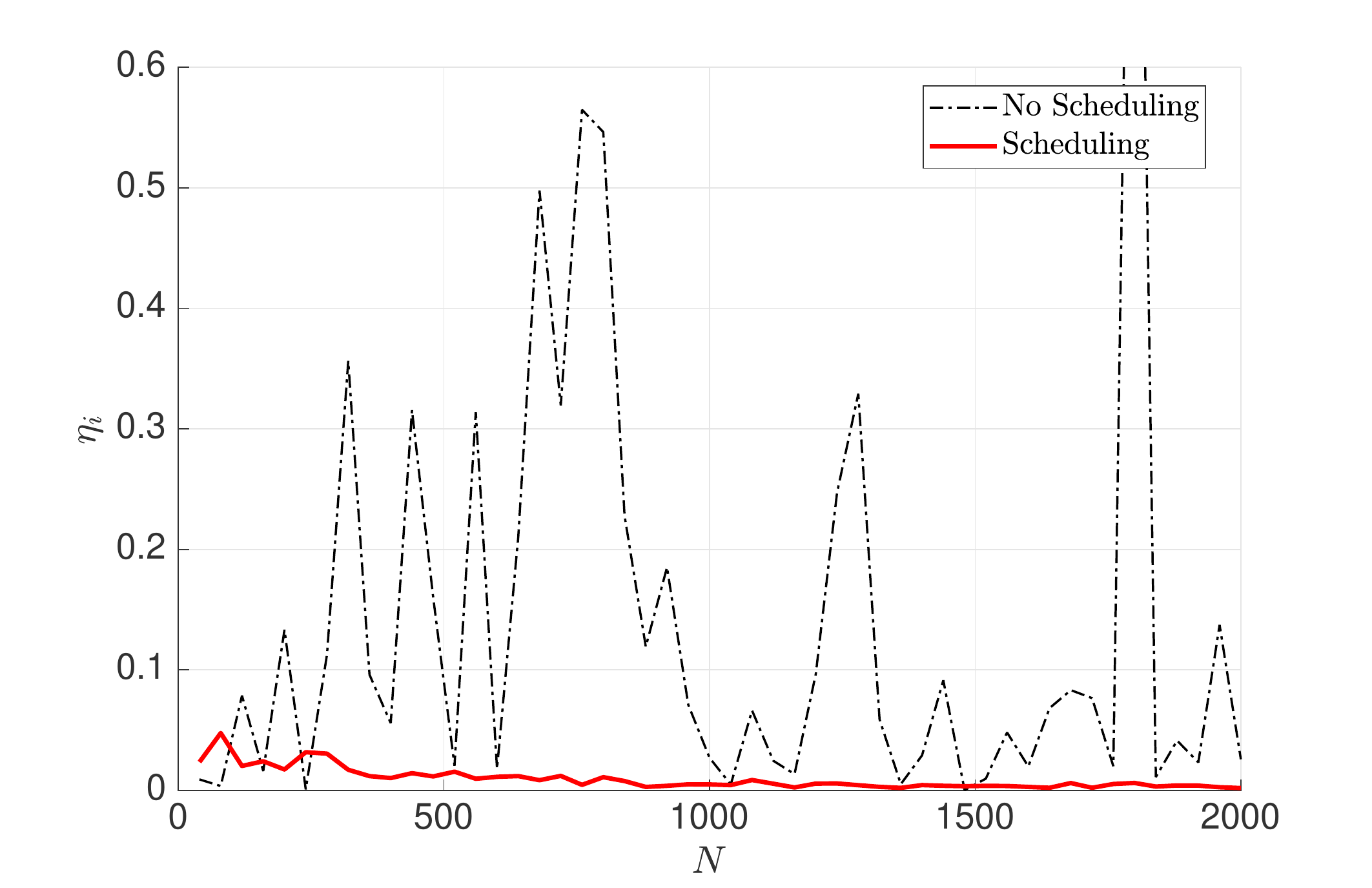}
	\caption{$\eta_i$ vs $N$ with and without scheduling.}
	\label{scheduling}	
\end{figure}
In Fig.~\ref{scheduling}, we show the instantaneous behavior of $\eta_i$ that causes  $\mathbb{E}[\eta_i]$ to grow. A simple channel is assumed with 20 paths, half-wavelength antenna spacing in a ULA, unit power rays, $\alpha=10$ and all rays have a $U[0,2\pi]$ distribution. With no control over the users entering the system, occasional large interference values occur as interfering user rays nearly align with the desired user. This is observed by the spikes in the curve labelled "no scheduling." Note that these spikes keep occurring even for massive antenna numbers, up to 2000. In contrast, we show  the trivial scheduling scheme introduced in Sec.~\ref{sched_ULA} with a protection target of $\epsilon=0.1$, equivalent to an angular separation of $0.57^\circ$. Here, the value of $\eta_i$ is well-behaved and decays to zero. 
\vspace*{-0.3cm}
\section{Conclusion}
The fundamental properties of massive MIMO have been identified with great generality for a broad class of ray-based models with a ULA or a UPA at the BS. The generality and insight possible is considerably greater than can be achieved with statistical channel models. In particular, we show that CH may or may not occur depending on the model used and FP is guaranteed for all continuous angular distributions. Although LSP will not normally hold, as the mean interference grows logarithmically relative to the desired channel, the implications for massive MIMO are excellent. As the number of users grows, the mean interference does grow relative to the desired channel but extremely slowly and this is further reduced by practical considerations, such as the attenuation of end-fire radiation caused by typical array patterns.   In addition, we prove that this mean interference growth can be avoided by trivial scheduling schemes.
\vspace*{-0.4cm}
\begin{appendices}\label{app}
\section{Proof of Theorem 1} \label{proofThm1}	
We note that 
\begin{align} \label{eq:expectation}
\mathbb{E}[e^{-jq\hat\phi_{ir}}]&=\int_{-2\pi{d}}^{2\pi{d}} e^{-jqx}f_{\hat\phi}(x)dx,
\end{align}
where $f_{\hat\phi}(\cdot)$ is PDF of $\hat\phi_{ir}$. Now, $\hat\phi_{ir}=2\pi{d}\text{sin}\phi_{ir}$ is a non one-to-one transformation of $\phi_{ir}$. Using standard transformation theory, 
 we obtain
 \small{
\begin{align} 
f_{\hat\phi}(x)=\dfrac{p(x)}{\sqrt{l^2-x^2}}, \ -l \leq x \leq l,
\end{align}
}\normalsize
where $l=2\pi{d}$ and
\begin{align} \label{eq:expectation2}
p(x)=\begin{cases}
f_{\phi}\left(\text{sin}^{-1}\left(\frac{x}{l}\right)\right)+f_{\phi}\left(\pi-\text{sin}^{-1}\left(\frac{x}{l}\right)\right), \ \  x \geq 0\\
f_{\phi}\left(\text{sin}^{-1}\left(\frac{x}{l}\right)\right)+f_{\phi}\left(-\pi-\text{sin}^{-1}\left(\frac{x}{l}\right)\right), x < 0.
\end{cases}
\end{align}

Hence, (\ref{eq:expectation}) is rewritten as
\begin{align} \label{eq:expP1}
\mathbb{E}[e^{-jq\hat\phi_{ir}}]&=\int_{-l}^{l} e^{-jqx}\dfrac{p(x)}{\sqrt{l^2-x^2}}dx.
\end{align}
Using the notation in \cite[Eq.1, p.~15]{lighthill1958introduction}, the Fourier transform (FT) of a function $f(x)$ is
\begin{align} 
g(y)=\int_{-\infty}^{\infty}f(x)e^{-{j}2\pi{x}y}dx.
\end{align}
If we set $y=\dfrac{q}{2{\pi}}$, then
\begin{align}  \label{eq:FT}
g\left(\dfrac{q}{2{\pi}}\right)=\int_{-\infty}^{\infty}f(x)e^{-jqx}dx.
\end{align}
Using the Heaviside function, $H(x)$, we can write (\ref{eq:expP1}) as a FT in the same format as (\ref{eq:FT}), 
\begin{align} \label{eq:expP}
\mathbb{E}[e^{-jq\hat\phi_{ir}}]&=\int_{-\infty}^{\infty} e^{-jqx}\left(H(x+l)-H(x-l)\right)\dfrac{p(x)}{\sqrt{l^2-x^2}}dx. 
\end{align}   
Hence, defining $f(x)=\left(H(x+l)-H(x-l)\right)\dfrac{p(x)}{\sqrt{l^2-x^2}}$, allows $\mathbb{E}[e^{-jq\hat\phi_{ir}}]$ to be computed as the FT of $f(x)$.
\par
This formulation is particularly useful as we can now leverage known results on the asymptotics of FTs as $q \rightarrow \infty$ \cite{lighthill1958introduction}. These results depend on the singularities of $f(x)$ so we first discuss the nature of these singularities. Clearly, $f(x)$
has singularities at $x=\pm{l}$ and at any singularities of $p(x)$. Note that the singularities at $x=\pm{l}$ are infinite discontinuities (indicating that the value of $f(x)$ will grow infinitely large as $x$ approaches $\pm{l}$). In contrast, the singularities of $p(x)$ are never infinite discontinuities for any proposed, practical angular distribution models. 
Models such as the wrapped Gaussian have no singularities inside $(-l,l)$ while the  Laplacian has only a non-differentiable point at the peak. Hence, the singularities at $x=\pm{l}$ are the worst. 
The general principle presented in  \cite[p.~55]{lighthill1958introduction} is that the 'worst' singularity\footnote{The singularity $x=x_m$ of a function, $f(x)$, is worst if $f(x)$ is of order $|x-x_m|^{\beta}$ near $x_m$ and $\beta$ is the smallest value for all singularities\cite[p.~55]{lighthill1958introduction}.} of a function contributes the leading term to the asymptotic expression for its FT. Thus in our case, we only need to consider the two singularities at $x=\pm{l}$. 
Near $x_{1}=-l$, $f(x)$ behaves like  $F_{1}(x)={H(x+l)p(-l)}({{2l}(l+x)})^{-1/2}$ and similarly near $x_{2}=l$, $f(x)$ behaves like $F_{2}(x)={(1-H(x-l))p(l)}({2l}(l-x))^{-1/2}$. Rewriting, we obtain
\begin{align} \label{eq:F1}
F_{1}(x)=\dfrac{H(x+l)p(-l)}{\sqrt{2l}}|x+l|^{-\frac{1}{2}},
\end{align}
\begin{align} \label{eq:F2}
F_{2}(x)=\dfrac{p(l)}{\sqrt{2l}}|x-l|^{-\frac{1}{2}}-\dfrac{H(x-l)p(l)}{\sqrt{2l}}|x-l|^{-\frac{1}{2}}.
\end{align}
From \cite[Theorem 19, p.~52]{lighthill1958introduction}, we know that if a generalised function, $f(x)$, has a finite number of singularities at \{$x=x_{1}, x_{2}, x_{3},..., x_{m}$\}, and for each of them $f(x)-F_{m}(x)$ has absolutely integrable $N^\text{th}$ order derivatives in an interval including $x_{m}$, where $F_{m}(x)$ is a linear combination of functions of type $|x-x_{m}|^{\beta}$, $|x-x_{m}|^{\beta}\text{sgn}(x-x_{m})$, $|x-x_{m}|^{\beta}\text{log}|x-x_{m}|$, $|x-x_{m}|^{\beta}\text{log}|x-x_{m}|\text{sgn}(x-x_{m})$, and if $f^{(N)}(x)$ is well behaved at infinity, then $g(y)$, the FT of $f(x)$, satisfies $g(y)=\sum_{m=1}^{M}G_{m}(y)+o(|y|^{-N})$, as $|y|\longrightarrow \infty$, where $G_{m}(y)$ is the FT of $F_{m}(x)$. Using this, 
\vspace*{-0.3cm}
\begin{align} 
g\left(\dfrac{q}{2\pi}\right) \sim G_{1}\left(\dfrac{q}{2\pi}\right)+ G_{2}\left(\dfrac{q}{2\pi}\right),
\end{align}
where $G_{1}$ and $G_2$ are the FTs of $F_1(x)$ and $F_2(x)$ in (\ref{eq:F1}) and (\ref{eq:F2}) and $\sim$ denotes asymptotic equivalence defined in \cite[p.~15]{abramowitz1964handbook}. From \cite[Table 1, p.~43]{lighthill1958introduction}, the FTs required are
\begin{align} \label{eq:table1} \nonumber
\mathcal{F}(|x-l|^{-\frac{1}{2}})&= e^{-2\pi{jl}y}|y|^{-\frac{1}{2}}, \\
\mathcal{F} (H(x+l)|x+l|^{-\frac{1}{2}})&=e^{2\pi{jl}y-\frac{1}{4}j\pi{ \text{sgn}(y)}}|2y|^{-\frac{1}{2}}, \\  \nonumber
\mathcal{F}(H(x-l)|x-l|^{-\frac{1}{2}})&= e^{-2\pi{jl}y-\frac{1}{4}j\pi{ \text{sgn}(y)}}|2y|^{-\frac{1}{2}}.
\end{align}
Using (\ref{eq:table1}), we obtain
\small{
\begin{align}  \label{eq:finalOR}
&g\left(\dfrac{q}{2\pi}\right) 
\sim G_{1}\left(\dfrac{q}{2\pi}\right)+ G_{2}\left(\dfrac{q}{2\pi}\right) 
=\sqrt{\dfrac{\pi}{lq}}\left(\dfrac{p(-l)}{\sqrt{2}}e^{j(lq-\frac{\pi}{4})}+{p(l)}e^{-jlq}-\dfrac{p(l)}{\sqrt{2}}\sqrt{\frac{\pi}{q}}e^{-j(lq+\frac{\pi}{4})}\right).
\end{align}
}\normalsize
Substituting $p(l)=2f_{\phi}(\frac{\pi}{2})$, $p(-l)=2f_{\phi}(\frac{-\pi}{2})$ and $l=2\pi{d}$ into (\ref{eq:finalOR}), and after some simplification we obtain the result in Theorem 1.
\vspace*{-0.7cm}
\section{Proof of Theorem 2} \label{proofThm2}	
	We drop the subscripts to re-express $M_{v,w}$ as follows,
	\small{
	\begin{align}  \nonumber
	M&=\mathbb{E}_\theta\left[\mathbb{E}_\phi[e^{-j2\pi\nu{\text{sin}}\theta\text{sin}{\tilde\phi}}]\right],
	\end{align} 
}\normalsize
	where $\tilde\phi=\phi-\Delta+\frac{\pi}{2}$.
	We get
		\small{
	\begin{align}  \label{eq: M}
	M&=\int_{-\frac{\pi}{2}}^{0}\left\{\mathbb{E}_\phi[e^{-j2\pi\nu|{\text{sin}}\theta|\text{sin}{\tilde{\tilde\phi}}}]\right\}f(\theta)d\theta  +\int_{0}^{\frac{\pi}{2}}\left\{\mathbb{E}_\phi[e^{-j2\pi\nu|{\text{sin}}(\theta)|\text{sin}{{\tilde\phi}}}]\right\}f(\theta)d\theta,
	\end{align} 
}\normalsize
	where ${\tilde{\tilde\phi}}=-{{\tilde\phi}}$. Setting $q=\nu$, $d=|\text{sin}\theta|$, allows us to use {Theorem 1} to give the asymptotic version of the expected values in (\ref{eq: M}). Hence,
	\vspace*{-0.3cm}
	\small{
	\begin{align}   \label{UPA:M1}
	&M\sim \int_{-\frac{\pi}{2}}^{0}\dfrac{f(\theta)}{\sqrt{\nu{|\text{sin}\theta|}}}\left\{f_{\tilde{\tilde\phi}}\left(-\frac{\pi}{2}\right)e^{j\left(2\pi\nu|\text{sin}\theta|-\frac{\pi}{4}\right)}\right\}d\theta 
	+\int_{-\frac{\pi}{2}}^{0}\dfrac{f(\theta)}{\sqrt{\nu{|\text{sin}\theta|}}}\left\{f_{\tilde{\tilde\phi}}\left(\frac{\pi}{2}\right)e^{-j\left(2\pi\nu|\text{sin}\theta|-\frac{\pi}{4}\right)}\right\}d\theta \\\nonumber
	\vspace*{-0.3cm}
	&+ \int_{0}^{\frac{\pi}{2}}\dfrac{f(\theta)}{\sqrt{\nu{|\text{sin}\theta|}}}\left\{f_{{\tilde\phi}}\left(-\frac{\pi}{2}\right)e^{j\left(2\pi\nu|\text{sin}\theta|-\frac{\pi}{4}\right)}\right\}d\theta 
	+\int_{0}^{\frac{\pi}{2}}\dfrac{f(\theta)}{\sqrt{\nu{|\text{sin}\theta|}}}\left\{f_{{\tilde\phi}}\left(\frac{\pi}{2}\right)e^{-j\left(2\pi\nu|\text{sin}\theta|-\frac{\pi}{4}\right)}\right\}d\theta. \\ \nonumber
	\end{align}
	\vspace*{-0.1\columnwidth}	
}\normalsize 
	
Substituting $f_{\tilde{\tilde\phi}}\left(-\frac{\pi}{2}\right)=f_{\phi}(\Delta)$, $f_{\tilde{\tilde\phi}}\left(\frac{\pi}{2}\right)=f_{\phi}(\Delta-\pi)$, $f_{{\tilde\phi}}\left(-\frac{\pi}{2}\right)=f_{\phi}(\Delta-\pi)$, and $f_{{\tilde\phi}}\left(\frac{\pi}{2}\right)=f_{\phi}(\Delta)$ into (\ref{UPA:M1}), we have
	\begin{align}   \label{UPA:M2}
	M&\sim \dfrac{f_\phi(\Delta)e^{-j\frac{\pi}{4}}}{\sqrt{\nu}}I_1+\dfrac{f_\phi(\Delta-\pi)e^{j\frac{\pi}{4}}}{\sqrt{\nu}}I_2 
	+\dfrac{f_\phi(\Delta-\pi)e^{-j\frac{\pi}{4}}}{\sqrt{\nu}}I_3+\dfrac{f_\phi(\Delta)e^{j\frac{\pi}{4}}}{\sqrt{\nu}}I_4,
	\end{align} 
	where $I_1^{*}=I_2=\displaystyle\int_{-\frac{\pi}{2}}^{0}\dfrac{f(\theta)e^{j2\pi\nu{\text{sin}\theta}}}{\sqrt{|\text{sin}\theta|}}d\theta$ and $I_3^{*}=I_4=\displaystyle\int_{0}^{\frac{\pi}{2}}\dfrac{f(\theta)e^{-j2\pi\nu{\text{sin}\theta}}}{\sqrt{\text{sin}\theta}}d\theta$. Hence, we only need to compute $I=\displaystyle\int_{0}^{\frac{\pi}{2}}\dfrac{g(\theta)e^{-j2\pi\nu{\text{sin}\theta}}}{\sqrt{\text{sin}\theta}}d\theta$, where $g(\theta)=f(\theta)$ for $I_3$ and $I_4$ and $g(\theta)=f(-\theta)$ for $I_1$ and $I_2$. Let $X=\text{sin}\theta$ and rewrite $I$ as
\small{
	\begin{align} I&=\displaystyle\int_{0}^{1}\dfrac{g(\text{sin}^{-1}(x))e^{-j2\pi\nu{x}}}{\sqrt{x}\sqrt{1-x^2}}dx
	=\displaystyle\int_{0}^{1}e^{-j2\pi\nu{x}}q(x)dx,
	\end{align}
}\normalsize
	where $q(x)=\dfrac{g(\text{sin}^{-1}(x))}{\sqrt{x}\sqrt{1-x^2}}\left[H(x)-H(x-1)\right]$. As in {Theorem 1}, $q(x)$ has its worst singularities at $x \in \{0, 1\}$. Using \cite[p.~55]{lighthill1958introduction} again, we have
	$q(x)\sim F_1(x)=g(0)H(x)|x|^{-\frac{1}{2}}$, when $x\longrightarrow0$, and $q(x)\sim F_2(x)=-\frac{1}{\sqrt{2}}g(\frac{\pi}{2})|x-1|^{-\frac{1}{2}}+\frac{1}{\sqrt{2}}g(\frac{\pi}{2})|x-1|^{-\frac{1}{2}}H(x-1)$, when $x\longrightarrow1$.
	Thus, according to  \cite[{Theorem 19}, p.~52]{lighthill1958introduction}, we have $I \sim G_1(y)+G_2(y)$, where
	$G_i(\cdot)$ is the Fourier transform of $F_i(\cdot)$ and $y=\nu/{2\pi}$. Hence,

	\begin{align} \label{eq: M2}
	I &\sim g(0)\dfrac{e^{-j\frac{\pi}{4}\text{sgn}(y)}}{\sqrt{2}}|y|^{-\frac{1}{2}} 
	-g\left(\frac{\pi}{2}\right)\dfrac{e^{-j2\pi{y}}}{\sqrt{2}}|y|^{-\frac{1}{2}} 
	+g\left(\frac{\pi}{2}\right)e^{-j2\pi{y}}\dfrac{e^{-j\frac{\pi}{4}\text{sgn}(y)}}{{2}}|y|^{-\frac{1}{2}}.
	\end{align}
	Substituting $\nu=2\pi{y}$ in (\ref{eq: M2})  and nothing that $\nu>0$ gives

	\begin{align} \label{eq:I}
	I &\sim  \dfrac{g(0)e^{-j\frac{\pi}{4}}}{\sqrt{2\nu}}-\dfrac{g(\frac{\pi}{2})e^{-j2\pi{\nu}}}{\sqrt{2\nu}}+\dfrac{g(\frac{\pi}{2})e^{-j2\pi{\nu}}e^{-j\frac{\pi}{4}}}{2\sqrt{\nu}}
	=\frac{1}{\sqrt{\nu}}\left\{g(0)e^{-j\frac{\pi}{4}}-\dfrac{g(\frac{\pi}{2})e^{-j2\pi{\nu}+j\frac{\pi}{4}}}{2}\right\}.
	\end{align}

Substituting (\ref{eq:I}) into (\ref{UPA:M2}) we get
\small{
	\begin{align} \label{UPA:M3} \nonumber
	M &\sim  \dfrac{f_\phi(\Delta)e^{-j\frac{\pi}{4}}}{\nu}\left\{\dfrac{f_\theta(0)e^{j\frac{\pi}{4}}}{\sqrt{2}}-\dfrac{f_\theta(-\frac{\pi}{2})}{2}e^{j(2\pi{\nu}-\frac{\pi}{4})}\right\} 
	+\dfrac{f_\phi(\Delta-\pi)e^{j\frac{\pi}{4}}}{\nu}\left\{\dfrac{f_\theta(0)e^{-j\frac{\pi}{4}}}{\sqrt{2}}-\dfrac{f_\theta(-\frac{\pi}{2})e^{-j(2\pi{\nu}-\frac{\pi}{4})}}{2}\right\}, \\ \nonumber
	+&\dfrac{f_\phi(\Delta-\pi)e^{-j\frac{\pi}{4}}}{\nu}\left\{\dfrac{f_\theta(0)e^{j\frac{\pi}{4}}}{\sqrt{2}}-\dfrac{f_\theta(\frac{\pi}{2})e^{j(2\pi{\nu}-\frac{\pi}{4})}}{2}\right\} 
	+\dfrac{f_\phi(\Delta)e^{j\frac{\pi}{4}}}{\nu}\left\{\dfrac{f_\theta(0)e^{-j\frac{\pi}{4}}}{\sqrt{2}}-\dfrac{f_\theta(\frac{\pi}{2})}{2}e^{-j(2\pi{\nu}-\frac{\pi}{4})}\right\}, \\ \nonumber
	&=\dfrac{j}{2\nu}\left\{-f_{\phi}(\Delta-\pi)f_{\theta}\left(-\tfrac{\pi}{2}\right)e^{-j2\pi\nu}\right\} 
	+\dfrac{j}{2\nu}\left\{f_{\phi}(\Delta-\pi)f_{\theta}\left(\tfrac{\pi}{2}\right)e^{j2\pi\nu}\right\} 
	+\dfrac{j}{2\nu}\left\{f_{\phi}(\Delta)f_{\theta}\left(-\tfrac{\pi}{2}\right)e^{j2\pi\nu}\right\}, \\ 
	&-\dfrac{j}{2\nu}\left\{f_{\phi}(\Delta)f_{\theta}\left(\tfrac{\pi}{2}\right)e^{-j2\pi\nu}\right\} 
	+\dfrac{\sqrt{2}f_{\theta}(0)}{\nu}(f_{\phi}(\Delta)+f_{\phi}(\Delta-\pi)) \triangleq\dfrac{\alpha(\nu,\Delta)}{\nu}.
	\end{align} 
}\normalsize
	According to a two-dimensional version of the integral test  (a one-dimensional version of the integral test can be found in \cite[Theorem 9.10, p.~619]{larson2009calculus}), (\ref{eq:nu_{v,w}}) converges if and only if the following expression converges as $N_x\rightarrow\infty$, $N_y\rightarrow \infty$, 
	\small{
	\begin{align}  \label{UPA:M5}
	Z_{0}=\displaystyle\int_{-N_x}^{N_x}\displaystyle\int_{y \in \mathcal{Y}}\left(1-\dfrac{|x|}{N_x}\right)\left(1-\dfrac{|y|}{N_y}\right)|M_{x,y}|^2dydx.
	\end{align} 
}\normalsize
	where $\mathcal{Y}=(-N_y,-1) \cup (1,N_y)$. Note that the interval$[-1,+1]$ has been cut out of the integration zone for $y$. This is valid as it simply reflects that the first few terms indexed by $w$ in (\ref{eq:nu_{v,w}}) are finite. As $M_{x,y}\sim \dfrac{\alpha(\nu,\Delta)}{\nu}$, we have
	\begin{align}  \label{UPA:M6}
	Z_{0}\sim \displaystyle\int_{-N_x}^{N_x}\displaystyle\int_{y \in \mathcal{Y}}\left(1-\dfrac{|x|}{N_x}\right)\left(1-\dfrac{|y|}{N_y}\right)\dfrac{|\alpha(\nu,\Delta)|^2}{x^2d^2_x+y^2d^2_y}dydx.
	\end{align}
	Since $|\alpha(\nu,\Delta)|^2$ is bounded and non-vanishing, the limit of $Z_{0}$ exists if and only if the limit of $Z_1$ exists, where 
	\begin{align} \nonumber 
	Z_{1}&\sim \displaystyle\int_{-N_x}^{N_x}\displaystyle\int_{y \in \mathcal{Y}}\left(1-\dfrac{|x|}{N_x}\right)\left(1-\dfrac{|y|}{N_y}\right)\dfrac{1}{x^2+y^2}dydx,\\ 
	&=4\displaystyle\int_{0}^{N_x}\displaystyle\int_{1}^{N_y}\left(1-\dfrac{x}{N_x}\right)\left(1-\dfrac{y}{N_y}\right)\dfrac{1}{x^2+y^2}dydx
	=I_A+I_B+I_C+I_D,
	\end{align}
	where $I_A=\displaystyle\int_{0}^{N_x}\displaystyle\int_{1}^{N_y}\dfrac{1}{x^2+y^2}dydx$, $I_B=-\displaystyle\int_{0}^{N_x}\displaystyle\int_{1}^{N_y}\frac{1}{N_x}\dfrac{x}{x^2+y^2}dydx$,\\ $I_C=\displaystyle\int_{0}^{N_x}\displaystyle\int_{1}^{N_y}\frac{1}{N_xN_y}\dfrac{xy}{x^2+y^2}dydx$, and $I_D=-\displaystyle\int_{0}^{N_x}\displaystyle\int_{1}^{N_y}\frac{1}{N_y}\dfrac{y}{x^2+y^2}dydx$.
	Using the integral result $\int_{0}^{\infty}\left(x^2+y^2\right)^{-1}dx=\pi/{2y}$ \cite[Eq. 2.124.1, p.~71]{gradshteyn2014table}, we are able to investigate the limits 	\vspace*{-0.3cm} of $I_A, I_B, I_C$ and $I_D$. First, we  consider $I_D$:
	\begin{align}   \nonumber
	\lim\limits_{N_x\rightarrow\infty, N_y\rightarrow\infty}I_D&=-\lim\limits_{ N_y\rightarrow\infty}\displaystyle\int_{1}^{N_y}\frac{y}{N_y}\displaystyle\int_{0}^{\infty}\dfrac{y}{x^2+y^2}dxdy
	=-\lim\limits_{ N_y\rightarrow\infty}\displaystyle\int_{1}^{N_y}\dfrac{\pi}{2N_y}dy=-\frac{\pi}{2},
	\end{align}
	which is finite. Similarly, $I_B$ and $I_C$ are also finite in the limit. Thus, the limiting behavior of (\ref{eq:nu_{v,w}}) depends on $I_A$, and we  have
	\begin{align}   \nonumber
	\lim\limits_{N_x\rightarrow\infty, N_y\rightarrow\infty}I_A&=\lim\limits_{ N_y\rightarrow\infty}\displaystyle\int_{1}^{N_y}\displaystyle\int_{0}^{\infty}\dfrac{1}{x^2+y^2}dxdy 
	=\lim\limits_{ N_y\rightarrow\infty}\displaystyle\int_{1}^{N_y}\dfrac{\pi}{2y}dy
	=\lim\limits_{ N_y\rightarrow\infty}\frac{\pi}{2}\text{log}(N_y).
	\end{align}
	Hence, $I_A$ grows logarithmically and (\ref{eq:nu_{v,w}}) grows logarithmically as desired.	
\vspace*{-0.1cm}
\section{Derivation of $\mu_{\text{UPA}}$ for Uniform Angular Distributions} \label{mu_for_UPA}	
In this scenario, beginning with the basic definition of $\mu_{\text{UPA}}$, we have
\begin{align} \label{eq:mu_UPA}\nonumber
\mu_{\text{UPA}}&=\frac{1}{N}\mathbb{E}\left[\left|\ba_{ir}^{\text{H}}\ba_{js}\right|^2\right] 
=\frac{1}{N}\sum_{k=1}^{N}\sum_{l=1}^{N}\mathbb{E}\left[a_{irk}^*a_{jsk}a_{jsl}^*a_{irl}\right]
=\frac{1}{N}\sum_{k=1}^{N}\sum_{l=1}^{N}\mathbb{E}\left[a_{irk}^*a_{irl}\right]\mathbb{E}\left[a_{jsl}^*a_{jsk}\right], \\ 
&=\frac{1}{N}\sum_{k=1}^{N}\sum_{l=1}^{N}\left|\mathbb{E}\left[a_{irk}^*a_{irl}\right]\right|^2
=\frac{1}{N}\sum_{k_x=1}^{N_x}\sum_{k_y=1}^{N_y}\sum_{l_x=1}^{l_x}\sum_{l_y=1}^{N_y}\left|\mathbb{E}\left[a_{irxk_x}^{*}a_{iryk_y}^{*}a_{irxl_x}a_{iryl_y}\right]\right|^2. 
\vspace*{-0.8cm}
\end{align}	
The expectation in (\ref{eq:mu_UPA}) is given by
\begin{align} 
I_1
&=\mathbb{E}[\text{exp}\{j2\pi{\text{sin}\theta}(d_y(k_y-l_y){\text{sin}\phi}+d_x(k_x-l_x){\text{cos}\phi})\}], \\ \nonumber
\end{align}
\vspace*{-2cm}
Taking expectation over $\phi$ first gives an integral of the form
\begin{align}
&\dfrac{1}{2\pi}\int_{0}^{2\pi}\text{exp}\{jz(\alpha\text{sin}\phi+\beta\text{cos}\phi)\}d{\phi}=\dfrac{1}{2\pi}\int_{0}^{2\pi}\text{exp}\{jz(\sqrt{\alpha^2+\beta^2}\text{sin}(\phi+\Delta))\}d{\phi}=J_0(z\sqrt{\alpha^2+\beta^2}),
\end{align}
where $\Delta=\text{tan}^{-1}(\beta/\alpha)$. Hence,
\begin{align} 
\small
I_1&=\mathbb{E}\left[J_0\left(2\pi{\text{sin}\theta\sqrt{d_y^2(k_y-l_y)^2+d_x^2(k_x-l_x)^2}}\right)\right]=\frac{1}{\pi}\int_{0}^{\pi}J_0(\chi{\text{sin}\theta})d\theta,
\end{align}
where  $\chi=2\pi{\sqrt{d_y^2(k_y-l_y)^2+d_x^2(k_x-l_x)^2}}$.

From \cite[Eq.6, p.~724]{gradshteyn2014table}, we have
\begin{align}
\int_{0}^{\pi}J_0(2z\text{sin}x)\text{cos}(2nx)dx=\pi{J_n^2(x)}.
\end{align}
Thus
\begin{align}
\int_{0}^{\pi}J_0(2z\text{sin}x)dx=\pi{J_0^2(z)},
\end{align}
and we can rewrite $I_1$ as
\begin{align}
I_1 =J_0^2(\pi\sqrt{d_y^2(k_y-l_y)^2+d_x^2(k_x-l_x)^2}).
\end{align}
Then, substituting $I_1$ into  (\ref{eq:mu_UPA}) allows $\mu_{\text{UPA}}$ to be written as in  (\ref{eq:upa uniform}).
\vspace*{-0.5cm}
\section{Logarithmic growth of $\mu_{\text{UPA}}$ for Uniform Angular Distributions} \label{The logarithm growth of Uniform Distribution for UPA}
A simple change of indices, $m=r_y-s_y$, $n=r_x-s_x$ in (\ref{eq:upa uniform}) gives 
\small{
\begin{align} \label{C: mu upa} 
\mu_\text{UPA}=\sum_{m=1-N_y}^{N_y-1}\sum_{n=1-N_x}^{N_x-1}\left(1-\frac{|m|}{N_y}\right)\left(1-\frac{|n|}{N_x}\right)  J_0^4(A(m,n)),
\end{align}
}\normalsize
where $A(m,n)=\pi\sqrt{m^2d^2_y+n^2d^2_x}$. The sum in (\ref{C: mu upa}) is dominated by 
\begin{align} \label{C: mu upa1}
\mu_{1}=\sum_{m=1-N_y}^{N_y-1}\sum_{n=1-N_x}^{N_x-1}J_0^4(A(m,n)),
\end{align}
and it is easy to show that the remaining terms in (\ref{C: mu upa}) are finite as $N\rightarrow\infty$. Hence, the asymptotic behavior of $\mu_{UPA}$ is the same as for $\mu_1$. Similarly, the sum in (\ref{C: mu upa1}) is dominated by
\begin{align} \label{C: mu upa2}
\mu_{2}=4\sum_{m=1}^{N_y-1}\sum_{n=1}^{N_x-1}J_0^4(A(m,n)),
\end{align}
using the fact that $A(m,n)$ is an even function of $m$ and $n$ and neglecting finite terms. Again, the asymptotic behavior of $\mu_\text{UPA}$ is the same as for $\mu_2$. Using the asymptotic equivalence  \cite[Eq. 10.17.3]{olver2010nist}, $J_0(z)\sim \sqrt{\dfrac{2}{\pi{z}}}\text{cos}(z-\pi/4)$, we see that $\mu_\text{UPA}$ behaves like
\small{
 \begin{align} \label{C: mu upa3} 
 \mu_{3}&=16\sum_{m=1}^{N_y-1}\sum_{n=1}^{N_x-1}\dfrac{\text{cos}^4(A(m,n)-\pi/4)}{\pi^2A^2(m,n)}=\dfrac{16}{\pi^4}\sum_{m=1}^{N_y-1}\sum_{n=1}^{N_x-1}\dfrac{\text{cos}^4(A(m,n)-\pi/4)}{m^2d^2_y+n^2d^2_x}.
 \end{align}
}\normalsize
Application of a two-dimensional version of the integral test  (one-dimensional version of the integral test can be found in \cite[Theorem 9.10, p.~619]{larson2009calculus}) and some further analysis to handle the oscillations in the $\text{cos}^4(\cdot)$ function via upper and lower bounds shows that the asymptotic behavior of $\mu_3$ is the same as that of 
 \begin{align} \label{C: mu upa4} 
\mu_4&=\int_{1}^{N_y-1}\int_{1}^{N_x-1}\dfrac{1}{x^2d^2_x+y^2d^2_y}d_xd_y=\dfrac{1}{d_xd_y}\int_{d_y}^{(N_y-1)d_y}\int_{d_x}^{(N_x-1)d_x}\dfrac{1}{u^2+v^2}dudv.
\end{align}
Converting to polar coordinates, a simple upper bound on (\ref{C: mu upa4}) is 
 \begin{align} \label{C: mu upa5} 
\mu_4 &\leqslant \int_{0}^{\pi/2}\int_{\rho_\text{min}}^{\rho_\text{max}}\dfrac{1}{\rho^2}{\rho}d\rho{d\theta}=\frac{\pi}{2}\left(\text{log}(\rho_\text{max})-\text{log}(\rho_\text{min})\right),
\end{align}
where $\rho_\text{min}=\text{min}(d_x,d_y)$ and $\rho_\text{max}=\sqrt{2}\text{max}((N_x-1)d_x, (N_y-1)d_y)$. Hence, the upper limit on $\mu_4$ grows logarithmically with $N$ as $N^{1/2} \leqslant \text{max}(N_x, N_y)\leqslant N$, so that $\frac{1}{2}\text{log}N \leqslant \text{log}(\text{max}(N_x, N_y))\leqslant \text{log}N$. Similarly, when $N_x \rightarrow \infty$,  $N_y \rightarrow \infty$ as $N\rightarrow \infty$, $\mu_4$ can be lower bounded by a logarithmic function of $N$ by integrating over a sector of an annulus contained inside the integration region of (\ref{C: mu upa4}). Hence, $\mu_\text{UPA}$ grows logarithmically with $N$ as required.

\end{appendices}
%
%
%
\bibliographystyle{IEEEtran}
\bibliography{bibliography}



\end{document}